\documentclass[12pt]{article}

\usepackage{graphicx}
\usepackage{amsmath}
\usepackage{epsfig}

\newcommand\Tstrut{\rule{0pt}{4ex}}         % = `top' strut
\newcommand\Bstrut{\rule[-3ex]{0pt}{0pt}}   % = `bottom' strut

\def\square{\kern1pt\vbox{\hrule height 1.2pt\hbox{\vrule width 1.2pt\hskip 3pt
   \vbox{\vskip 6pt}\hskip 3pt\vrule width 0.6pt}\hrule height 0.6pt}\kern1pt}

\def \be{\begin{equation}}
\def \ee{\end{equation}}
\def \bea{\begin{eqnarray}}
\def \eea{\end{eqnarray}}
\def \th{\theta}
\def \del{\partial}
\def \a{\alpha}
\def \b{\beta}
\def \f{\frac}
\def \G{\Gamma}
\def \g{\gamma}

\def \lam{\lambda}

\def \si{\sigma}

\def \d{\delta}
\def \D{\Delta}
\def \e{\eta}
\def \th{\theta}
\def \l{\lambda}
\def \k{\kappa}

\begin{document}

\begin{titlepage}

\begin{flushright}
UFIFT-QG-13-02
\end{flushright}

\vspace{1cm}

\begin{center}
{\bf Graviton Corrections to Vacuum Polarization during Inflation}
\end{center}

\vspace{.5cm}

\begin{center}
Katie E. Leonard$^{\dagger}$ and R. P. Woodard$^{\ddagger}$
\end{center}

\vspace{.5cm}

\begin{center}
\it{Department of Physics \\
University of Florida \\
Gainesville, FL 32611}
\end{center}

\vspace{1cm}

\begin{center}
ABSTRACT
\end{center}
We use dimensional regularization to compute the one loop quantum
gravitational contribution to the vacuum polarization on de Sitter
background. Adding the appropriate BPHZ counterterms gives a fully
renormalized result which can be used to quantum correct Maxwell's
equations. We use the Hartree approximation to argue that the
electric field strengths of photons experience a secular suppression
during inflation.

\vspace{.5cm}

\begin{flushleft}
\end{flushleft}

\vspace{1cm}

\begin{flushleft}
$^{\dagger}$ e-mail: katie@phys.ufl.edu \\
$^{\ddagger}$ e-mail: woodard@phys.ufl.edu

\vspace{1cm} {\it Dedicated to Stanley Deser on the occasion of his
82nd birthday.}

\end{flushleft}
\end{titlepage}

\section{Introduction}\label{Intro}

Inflation produces a vast ensemble of infrared scalars and gravitons.
This is thought to be the source of primordial scalar and tensor
perturbations \cite{perts}. It is natural to wonder what effect these
ensembles have on other particles. That sort of question can be
answered by computing the scalar or graviton contribution to the
appropriate 1PI (one-particle-irreducible) 2-point function and then
using that result to quantum-correct the linearized field equation
for the particle in question.

The 1PI 2-point function for a scalar is known as its
``self-mass-squared'', $-i M^2(x;x')$, and the quantum-corrected,
linearized field equation for a massless, minimally coupled scalar is,
\begin{equation}
\partial_{\mu} \Bigl( \sqrt{-g} \, g^{\mu\nu} \partial_{\nu}
\varphi(x) \Bigr) - \int \!\! d^4x' M^2(x;x') \varphi(x') = 0 \; ,
\end{equation}
where $g_{\mu\nu}(x)$ is the spacelike metric tensor. The fermion's
1PI 2-point function is called its ``self-energy'',
$-i [\mbox{}_i \Sigma_j](x;x')$, and the quantum-corrected, linearized
field equation for a massless fermion is,
\begin{equation}
\sqrt{-g} \, e^{\mu}_{~a} \gamma^a_{ij} \Bigl( i\partial_{\mu}
\delta_{jk} \!-\! \frac12 A_{\mu b c} J^{bc}_{jk} \Bigr) \psi_k(x)
- \int \!\! d^4x' \Bigl[ \mbox{}_i \Sigma_j\Bigr](x;x') \psi_j(x') = 0 \; ,
\end{equation}
where $e_{\mu a}(x)$ is the vierbein field, $\gamma^a_{ij}$ are the
gamma matrices, $A_{\mu bc}(x)$ is the spin connection, and $J^{bc}
\equiv -\frac{i}4 [ \gamma^b , \gamma^c ]$ are the Lorentz generators.
The 1PI 2-point function for a photon has the evocative name
``vacuum polarization'', $+i [\mbox{}^{\mu} \Pi^{\nu}](x;x')$, and
the quantum-corrected Maxwell equation is,
\begin{equation}
\partial_{\nu} \Bigl( \sqrt{-g} \, g^{\nu\rho} g^{\mu\sigma}
F_{\rho\sigma}(x) \Bigr) + \int \!\! d^4x' \Bigl[\mbox{}^{\mu}
\Pi^{\nu} \Bigr](x;x') A_{\nu}(x') = J^{\mu}(x) \; , \label{QMax}
\end{equation}
where $F_{\mu\nu} \equiv \partial_{\mu} A_{\nu} - \partial_{\nu} A_{\mu}$
is the field strength tensor and $J^{\mu}(x)$ is the current density.
And the 1PI 2-point function for a graviton is termed the ``graviton
self-energy'', $-i[\mbox{}^{\mu\nu} \Sigma^{\rho\sigma}](x;x')$, and
the quantum-corrected, linearized Einstein equation is,
\begin{equation}
\sqrt{-g} \, \mathcal{D}^{\mu\nu\rho\sigma} h_{\rho\sigma}(x) -
\int \!\! d^4x' \Bigl[ \mbox{}^{\mu\nu} \Sigma^{\rho\sigma}
\Bigr](x;x') h_{\rho\sigma}(x') = \frac12 \kappa \sqrt{-g} \,
T^{\mu\nu}_{\rm lin}(x) \; ,
\end{equation}
where $\mathcal{D}^{\mu\nu\rho\sigma}$ is the Lichnerowicz operator,
$\kappa^2 \equiv 16 \pi G$ is the loop counting parameter of quantum
gravity and $T^{\mu\nu}_{\rm lin}$ is the linearized stress tensor.

Many results of this type have been derived in recent years, with
the background geometry of primordial inflation modeled using the
cosmological patch of de Sitter space. The effects of massless,
minimally coupled (MMC) scalars are simplest to study. A quartic
self-interaction leads MMC scalars to develop a growing mass
\cite{BOW}. The vacuum polarization from charged MMC scalars causes
the photon to develop a mass \cite{SQED} and engenders profound
changes in electrodynamic forces \cite{DW}. MMC scalars which are
Yukawa-coupled to a fermion make the fermion develop a growing mass
\cite{Yukawa}. And MMC scalars do not have any effect on gravitons
provided one can absorb certain surface terms into perturbative
corrections of the initial state \cite{SPW}.

The effects of inflationary gravitons are more difficult to work
out (everything is tougher in quantum gravity!), but studies have
been made of what they do to fermions and to MMC scalars. The
results are interestingly different: whereas inflationary gravitons
induce a slow growth of the fermion field strength \cite{MW1} they
have no secular effect on MMC scalars \cite{KW}. The difference seems
to be due to spin. A MMC scalar can only interact with gravitons
through its kinetic energy but this cannot mediate any secular
growth, in spite of the growing graviton field strength, because
the scalar's kinetic energy redshifts to zero exponentially fast.
By contrast, a fermion interacts with gravitons through its spin,
in addition to its kinetic energy, and the spin-spin interaction
remains effective even when the kinetic energy redshifts to zero
\cite{MW2}. The same thing seems to be true of a small mass
\cite{SPM}.

The importance of spin in mediating interactions between
inflationary gravitons and massless fermions suggests that there
might be comparably strong effects on other particles with spin such
as photons and gravitons. The graviton contribution to the one loop
graviton self-energy has been worked out \cite{TW1} but so far not
used to quantum-correct the linearized Einstein equation. The
purpose of this paper is to derive the one loop graviton
contribution to the vacuum polarization on de Sitter background. We
will use it to solve the quantum-corrected Maxwell equation in a
subsequent paper. (The flat space analog of this problem was carried
out as a warmup exercise \cite{LW}.) Our computation is done in
dimensional regularization, fully renormalized with the necessary
BPHZ (Bogoliubov-Parasiuk-Hepp-Zimmermann) counterterms \cite{BPHZ},
and reported in the noncovariant tensor basis whose efficacy has
been demonstrated in a recent study \cite{LPW1}. We also use the
Hartree approximation to argue that photons likely experience a
secular suppression of their electric field strengths.

This paper contains seven sections of which the first is this
Introduction. Section~\ref{Feynman} gives those of the Feynman rules
of Maxwell + Einstein which are needed for our computation. The
contribution from a single 4-point vertex is derived in
section~\ref{4point}. Section~\ref{3point} gives the much more
complicated contribution from two 3-point vertices. Renormalization
is accomplished in section~\ref{Renorm}. Although the use of our
result to quantum-correct Maxwell's equation is deferred to a latter
work, the Hartree approximation is employed in section~\ref{Hartree}
to argue that photons experience secular changes of the same
strength but opposite sign as those of fermions \cite{MW1,MW2}. Our
conclusions are given in section~\ref{Discuss}.

\section{Feynman Rules}\label{Feynman}

The purpose of this section is to present the formalism used to
compute the one loop quantum gravity contribution to the vacuum
polarization depicted in Fig.~\ref{photon}. We begin by describing
the background geometry. Then we use the primitive Lagrangians to
derive formal expressions for the first two diagrams of Fig.~\ref{photon}.
The longest subsection discusses our conventions for gauge fixing and
the resulting propagators. We next describe how the vacuum polarization
can be represented in terms of two structure functions. The section
closes by giving the counterterms needed for this computation.

\subsection{Our de Sitter Background}\label{dS}

We model primordial inflation as the cosmological patch of de
Sitter space. The invariant element is,
\begin{equation}
ds^2 = a^2 \left[-d\eta^2 + d\vec{x} \cdot d\vec{x} \right] \; ,
\end{equation}
where $a(\eta)=-\frac{1}{H\eta}=e^{Ht}$ is the scale factor and $H$
is the Hubble parameter. Whereas the spatial coordinates $\vec{x}$
take their usual values, the conformal time $\eta$ runs from
$\eta \rightarrow -\infty$ (the infinite past) to $\eta \rightarrow
0^{-}$ (the infinite future).

In representing functions such as propagators which depend upon two
points, $x^{\mu}$ and ${x'}^{\mu}$, we will make extensive use of
the de Sitter length function,
\begin{equation}
y(x;x') \equiv a(\eta) a(\eta') H^2 \left[|| \vec{x} - \vec{x}'||^2
- (| \eta - \eta'| - i \delta)^2 \right] \; .
\end{equation}
We also need the de Sitter breaking product of the scale factors
$a$ at $x^{\mu}$ and $a'$ at ${x'}^{\mu}$,
\begin{equation}
u \equiv \ln(aa') \; .
\end{equation}
Derivatives of $y$ and $u$ furnish a convenient basis for
representing bi-vector functions of $x^{\mu}$ and ${x'}^{\mu}$ such
as the vacuum polarization,
\begin{equation}
\partial_{\mu} y \quad , \quad \partial'_{\nu} y \quad , \quad
\partial_{\mu} \partial'_{\nu} y \quad , \quad \partial_{\mu} u
\quad , \quad \partial'_{\nu} u \; . \label{basis}
\end{equation}
It turns out that either taking covariant derivatives of any of the
five basis tensors (\ref{basis}), or contracting any two of them
into one another, produces metrics and more basis tensors
\cite{KW2,MTW2}.

\subsection{Our Primitive Diagrams}\label{Primitive}

\begin{figure}
\includegraphics[width=4.0cm,height=3.0cm]{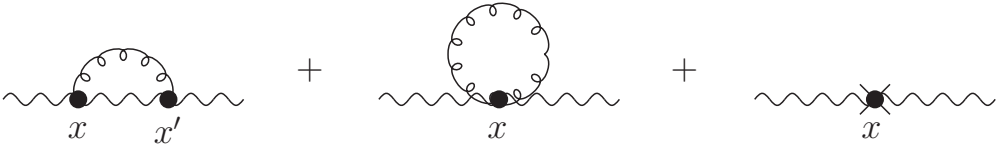}
\caption{Graviton contributions to the one loop vacuum polarization.
Photon propagators are wavy and graviton propagators are curly.}
\label{photon}
\end{figure}

The total Lagrangian consists of the primitive contributions
from general relativity and electromagnetism, plus the BPHZ
counterterms necessary for this computation,
\begin{equation}
\mathcal{L} = \mathcal{L}_{\rm GR} + \mathcal{L}_{\rm EM} +
\mathcal{L}_{\rm BPHZ} \; .
\end{equation}
The primitive Lagrangians of general relativity and electromagnetism
are,
\begin{equation}
\mathcal{L}_{\rm GR} = \frac1{16 \pi G} \, \Bigl( R \!-\! (D \!-\! 2)
\Lambda \Bigr) \sqrt{-g} \quad , \quad
\mathcal{L}_{\rm EM} = -\frac14 F_{\mu\nu} F_{\rho \sigma}
g^{\mu\rho} g^{\nu \sigma} \sqrt{-g} \; . \label{Lprimtive}
\end{equation}
The symbol $G$ stands for Newton's constant, while $\Lambda \equiv
(D-1) H^2$ is the cosmological constant. We employ a $D$-dimensional,
spacelike metric $g_{\mu\nu}$, with inverse $g^{\mu\nu}$ and
determinant $g = {\rm det}(g_{\mu\nu})$. Our affine connection and
Riemann tensor are,
\begin{eqnarray}
\Gamma^{\rho}_{~\mu\nu} & \equiv & \frac12 g^{\rho\sigma} \Bigl[
\partial_{\nu} g_{\sigma\mu} \!+\! \partial_{\mu} g_{\nu\sigma} \!-\!
\partial_{\sigma} g_{\mu\nu} \Bigr] \; , \\
R^{\rho}_{~\sigma\mu\nu} & \equiv & \partial_{\mu} \Gamma^{\rho}_{~\nu\sigma}
\!-\! \partial_{\nu} \Gamma^{\rho}_{~\mu\sigma} \!+\!
\Gamma^{\rho}_{~\mu\alpha} \Gamma^{\alpha}_{~\nu\sigma} \!-\!
\Gamma^{\rho}_{~\nu\alpha} \Gamma^{\alpha}_{~\mu\sigma} \; . \qquad
\end{eqnarray}
Our Ricci tensor is $R_{\mu\nu} \equiv R^{\rho}_{~\mu\rho\nu}$ and the
associated Ricci scalar is $R \equiv g^{\mu\nu} R_{\mu\nu}$. The
electromagnetic field strength tensor and its first covariant derivative
are,
\begin{equation}
F_{\mu\nu} \equiv \partial_{\mu} A_{\nu} \!-\! \partial_{\nu} A_{\mu}
\quad , \quad D_{\alpha} F_{\mu\nu} \equiv \partial_{\alpha}
F_{\mu\nu} \!-\! \Gamma^{\gamma}_{~\alpha\mu} F_{\gamma\nu} \!-\!
\Gamma^{\gamma}_{~\alpha\nu} F_{\mu\gamma} \; .
\end{equation}

We define the graviton field $h_{\mu\nu}(x)$ as the difference
between the full metric and its de Sitter background value
$a^2 \eta_{\mu\nu}$,
\begin{equation}
g_{\mu\nu}(x) \equiv a^2(\eta) \Bigl[ \eta_{\mu\nu} + \kappa
h_{\mu\nu}(x) \Bigr] \equiv a^2 \widetilde{g}_{\mu\nu}(x) \; ,
\end{equation}
where $\kappa^2 \equiv 16 \pi G$ is the loop counting parameter of
quantum gravity. We follow the usual conventions whereby a comma
denotes ordinary differentiation, the trace of the graviton field is
$h \equiv \eta^{\mu\nu} h_{\mu\nu}$, and graviton indices are raised
and lowered using the Minkowski metric, $h^{\mu}_{~\nu} \equiv
\eta^{\mu\rho} h_{\rho\nu}$ and $h^{\mu\nu} \equiv \eta^{\mu\rho}
\eta^{\nu\sigma} h_{\rho \sigma}$. Up to a surface term the
gravitational Lagrangian can be written as,
\begin{eqnarray}
\lefteqn{\mathcal{L}_{\rm GR} - {\rm Surface} = a^{D-2} \, \sqrt{-\widetilde{g}} \, \widetilde{g}^{\alpha\beta}
\widetilde{g}^{\rho\sigma} \widetilde{g}^{\mu\nu} } \nonumber \\
& & \hspace{2cm} \times \Biggl\{ \frac12 h_{\alpha\rho ,\mu}
h_{\nu\sigma ,\beta} \!-\! \frac12 h_{\alpha\beta ,\rho}
h_{\sigma\mu ,\nu} \!+\! \frac14 h_{\alpha\beta ,\rho}
h_{\mu\nu ,\sigma} \!-\! \frac14 h_{\alpha\rho ,\mu}
h_{\beta\sigma ,\nu} \Biggr\} . \qquad \label{Linv}
\end{eqnarray}
From Fig.~\ref{photon} one can see that we only need (\ref{Linv})
for the graviton propagator.

The only interactions we require descend from the second variational
derivative of the electromagnetic action,
\begin{equation}
\frac{\delta^2 S_{\rm EM}}{\delta A_{\mu}(x) \delta A_{\rho}(x') }
= \partial_{\kappa} \Biggl\{ \sqrt{-g(x)} \, \Bigl[ g^{\kappa\lambda}(x)
g^{\mu \rho}(x) \!-\! g^{\kappa\rho}(x) g^{\lambda\mu}(x) \Bigr]
\partial_{\lambda} \delta^D(x \!-\! x') \Biggr\} \; .
\end{equation}
The necessary vertex functions are obtained by expanding the metric
factors,
\begin{eqnarray}
\lefteqn{\sqrt{-g} \, \Bigl( g^{\kappa\lambda} g^{\mu\rho} \!-\!
g^{\kappa\rho} g^{\lambda\mu} \Bigr) \equiv a^{D-4} \Bigl(
\eta^{\kappa\lambda} \eta^{\mu\rho} \!-\! \eta^{\kappa\rho}
\eta^{\lambda\mu} \Bigr) } \nonumber \\
& & \hspace{2cm} + \kappa a^{D-4}
V^{\mu\rho\kappa\lambda\alpha\beta} h_{\alpha\beta} + \kappa^2
a^{D-4} U^{\mu\rho\kappa\lambda\alpha\beta\gamma\delta}
h_{\alpha\beta} h_{\gamma\delta} + O(\kappa^3) \; . \qquad
\end{eqnarray}
The tensor factors for the 3-point and 4-point vertices are,
\begin{eqnarray}
\lefteqn{V^{\mu\rho\kappa\lambda\alpha\beta} = \eta^{\alpha\beta}
\eta^{\kappa [\lambda} \eta^{\rho ] \mu} \!+\! 4 \eta^{\alpha) [\mu}
\eta^{\kappa ] [\rho} \eta^{\lambda ] (\beta} \; , } \label{Vvert} \\
\lefteqn{U^{\mu\rho\kappa\lambda\alpha\beta\gamma\delta} = \Bigl[\frac14
\eta^{\alpha\beta} \eta^{\gamma\delta} \!-\! \frac12 \eta^{\alpha (\gamma}
\eta^{\delta) \beta} \Bigr] \eta^{\kappa [\lambda} \eta^{\rho ] \mu} +
\eta^{\alpha\beta} \eta^{\gamma) [\mu} \eta^{\kappa] [\rho}
\eta^{\lambda] (\delta} } \nonumber \\
& & \hspace{-.5cm} + \eta^{\gamma\delta} \eta^{\alpha) [\mu} \eta^{\kappa]
[\rho} \eta^{\lambda] (\beta} \!+\! \eta^{\kappa (\alpha} \eta^{\beta)
[\lambda} \eta^{\rho ] (\gamma} \eta^{\delta) \mu} \!+\! \eta^{\kappa (\gamma}
\eta^{\delta) [\lambda} \eta^{\rho ] (\alpha} \eta^{\beta) \mu}
\!+\! \eta^{\kappa (\alpha} \eta^{\beta) (\gamma} \eta^{\delta) [\lambda}
\eta^{\rho ] \mu} \nonumber \\
& & \hspace{1.5cm} + \eta^{\kappa (\gamma} \eta^{\delta) (\alpha}
\eta^{\beta) [\lambda} \eta^{\rho ] \mu} + \eta^{\kappa [\lambda}
\eta^{\rho ] (\alpha} \eta^{\beta) (\gamma} \eta^{\delta) \mu}
+ \eta^{\kappa [\lambda} \eta^{\rho ] (\gamma} \eta^{\delta) (\alpha}
\eta^{\beta) \mu} \; . \qquad \label{Uvert}
\end{eqnarray}
Parenthesized indices are symmetrized and indices enclosed in square
brackets are anti-symmetrized.

If we call the graviton propagator $i[\mbox{}_{\alpha\beta}
\Delta_{\gamma\delta}](x;x')$ and the photon propagator
$i[\mbox{}_{\rho} \Delta_{\sigma}](x;x')$, we can give formal
expressions for the first two diagrams of Fig.~\ref{photon}. The one
constructed from a single 4-point vertex is,
\begin{equation}
i\Bigl[\mbox{}^{\mu} \Pi^{\nu}_{\rm 4pt}\Bigr](x;x') =
\partial_{\kappa} \Biggl\{ i\kappa^2 a^{D-4}
U^{\mu\nu\kappa\lambda\alpha\beta\gamma\delta}
\, i\Bigl[\mbox{}_{\alpha\beta} \Delta_{\gamma\delta} \Bigr](x;x)
\, \partial_{\lambda} \delta^D(x \!-\! x') \Biggr\} \; . \label{4pt1}
\end{equation}
The diagram constructed from two 3-point vertices is,
\begin{eqnarray}
\lefteqn{i\Bigl[\mbox{}^{\mu} \Pi^{\nu}_{\rm 3pt}\Bigr](x;x') =
\partial_{\kappa} \partial_{\theta}' \Biggl\{ i\kappa a^{D-4}
V^{\mu\rho\kappa\lambda\alpha\beta} \, i\Bigl[
\mbox{}_{\alpha\beta} \Delta_{\gamma\delta} \Bigr](x;x') } \nonumber \\
& & \hspace{5.5cm} \times i\kappa {a'}^{D-4}
V^{\nu\sigma\theta\phi\gamma\delta} \, \partial_{\lambda}
\partial_{\phi}' i\Bigl[\mbox{}_{\rho} \Delta_{\sigma}\Bigr](x;x')
\Biggr\} \; . \qquad \label{3pt1}
\end{eqnarray}

\subsection{Our Propagators}\label{props}

The quadratic part of the gravitational Lagrangian (\ref{Linv}) is,
\begin{equation}
\mathcal{L}^{(2)}_{\rm GR} = a^{D-2} \Biggl\{\frac12 h^{\rho\sigma , \mu}
h_{\mu \sigma , \rho} \!-\! \frac12 h^{\mu \nu}_{~~ ,\mu} h_{,\nu}
\!+\! \frac14 h^{,\mu} h_{, \mu} \!-\! \frac14 h^{\rho\sigma , \mu}
h_{\rho \sigma , \mu} \Biggr\} \; . \label{L2GR}
\end{equation}
Before fixing the gauge and giving the graviton propagator we must
digress to summarize the long and confusing debate between cosmologists
and mathematical physicists concerning the de Sitter invariance of free
gravitons \cite{HMM,MTW4}. Although the propagator equation can be made
de Sitter invariant by an appropriate choice of gauge, that does not
guarantee the de Sitter invariance of the solution. The classic
counter-example --- which plays an important role in our solution
for the graviton propagator --- is the propagator $i\Delta_A(x;x')$ of
a massless, minimally coupled scalar,
\begin{equation}
\partial_{\mu} \Bigl( \sqrt{-g} \, g^{\mu\nu} \partial_{\nu}
i\Delta_A(x;x') \Bigr) \equiv \sqrt{-g} \, \square i\Delta_A(x;x')
= i\delta^D(x \!-\! x') \; . \label{MMCeqn}
\end{equation}
Equation (\ref{MMCeqn}) is de Sitter invariant, but there is no de
Sitter invariant solution for $i\Delta_A(x;x')$ \cite{AF}. This can
be seen from the time dependence of the coincidence limit \cite{coinc},
\begin{equation}
i\Delta_A(x;x) = \Bigl({\rm Divergent\ Constant}\Bigr) +
\frac{H^2}{4 \pi^2} \times \ln(a) \; .
\end{equation}
If one chooses the ``E(3)'' vacuum \cite{BA} to preserve the spatial
homogeneity and isotropy of cosmology then the unique solution is \cite{OW},
\begin{eqnarray}
\lefteqn{i \Delta_A(x;x') =  i \Delta_{\rm cf}(x;x') } \nonumber \\
& & + \frac{H^{D-2}}{(4\pi)^{\frac{D}2}} \frac{\Gamma(D \!-\! 1)}{\Gamma(
\frac{D}2)} \left\{\! \frac{D}{D\!-\! 4} \frac{\Gamma^2(\frac{D}2)}{\Gamma(D
\!-\! 1)} \Bigl(\frac4{y}\Bigr)^{\frac{D}2 -2} \!\!\!\!\!\! - \pi
\cot\Bigl(\frac{\pi}2 D\Bigr) + \ln(a a') \!\right\} \nonumber \\
& & + \frac{H^{D-2}}{(4\pi)^{\frac{D}2}} \! \sum_{n=1}^{\infty}\! \left\{\!
\frac1{n} \frac{\Gamma(n \!+\! D \!-\! 1)}{\Gamma(n \!+\! \frac{D}2)}
\Bigl(\frac{y}4 \Bigr)^n \!\!\!\! - \frac1{n \!-\! \frac{D}2 \!+\! 2}
\frac{\Gamma(n \!+\!  \frac{D}2 \!+\! 1)}{\Gamma(n \!+\! 2)} \Bigl(\frac{y}4
\Bigr)^{n - \frac{D}2 +2} \!\right\} \! , \quad \label{DeltaA}
\end{eqnarray}
where $i \Delta_{\rm cf}(x;x')$ is the (de Sitter invariant) propagator
of a conformally coupled scalar,
\begin{equation}
{i\Delta}_{\rm cf}(x;x') = \frac{H^{D-2}}{(4\pi)^{\frac{D}2}} \Gamma\Bigl(
\frac{D}2 \!-\! 1\Bigr) \Bigl(\frac4{y}\Bigr)^{\frac{D}2-1} \; .
\end{equation}

The massless, minimally coupled scalar is especially relevant to
gravitons because Grishchuk showed that the physical components of
$h_{\mu\nu}$ (transverse, traceless and purely spatial) obey the
same equation \cite{Grishchuk}. Hence the graviton propagator must
break de Sitter invariance as well. A cosmologist would also see
this from the scale invariance of the tensor power spectrum
\cite{MTW4}. Mathematical physicists for years disputed that conclusion
because they found manifestly de Sitter invariant solutions to the
propagator equation which results from adding de Sitter invariant
gauge fixing functions to the quadratic Lagrangian (\ref{L2GR})
\cite{INVPROP}.

The discordant viewpoints have recently converged somewhat with the
demonstration that there is an obstacle to adding invariant gauge
fixing functions on any manifold, like de Sitter, which possesses
a linearization instability \cite{MTW1}. That still leaves open the
possibilities of either adding a noninvariant gauge fixing function
or else enforcing a de Sitter invariant gauge condition as a strong
operator equation. When either possibility is pursued, along with
the requirement that the resulting propagator can be expressed as a
superposition of plane wave mode functions \cite{TW2,MTW3,MTW5}, the
result is a de Sitter breaking solution of precisely the form implied
by the scale invariance of the tensor power spectrum \cite{Kleppe,KMW}.
However, when a de Sitter invariant gauge condition is employed,
in conjunction with analytic continuation techniques (either from
Euclidean de Sitter space, or in the mass-squared of certain scalar
propagators), the result is de Sitter invariant \cite{FH,IAM},
except for certain discrete choices of the gauge condition.

It is well known that analytic continuation fails to recover power
law infrared divergences \cite{KK,JMPW,MTW2}. In this context the
discrete problems, which have long been noted \cite{AM,Folacci}, seem
suspiciously like the special values at which a power law infrared
divergence --- which is always present --- happens to become
logarithmic and hence visible to an analytic continuation technique.
We will therefore employ a de Sitter breaking graviton propagator, which
seems to be the safer choice. Despite the continuing disagreements,
it is important to note that very little difference remains between
cosmologists and mathematical physicists. In particular, the various
de Sitter breaking solutions for the graviton propagator all give the
same result for the linearized Weyl-Weyl correlator \cite{PMW,MTW6},
and this agrees with the result from de Sitter invariant solutions
\cite{Kouris} once some errors are corrected \cite{Atsuchi}.

We fix the gauge by adding \cite{TW2},
\begin{equation}
\mathcal{L}_{\rm GRfix} = -\frac12 a^{D-2} \eta^{\mu\nu} F_{\mu} F_{\nu}
\; , \; F_{\mu} \equiv \eta^{\rho\sigma} \Bigl(h_{\mu\rho , \sigma}
\!-\! \frac12 h_{\rho \sigma , \mu} \!+\! (D\!-\!2) H a h_{\mu\rho}
\delta^0_{\sigma} \Bigr) . \label{GRfix}
\end{equation}
The resulting graviton propagator can be expressed as a sum of
constant tensor factors multiplied by scalar propagators \cite{TW2},
\begin{equation}
i \Bigl[\mbox{}_{\mu\nu} \Delta_{\rho\sigma}\Bigr](x;x') =
\sum_{I = A, B, C} \Bigl[\mbox{}_{\mu\nu} T^I_{\rho\sigma}\Bigr]
\times i\Delta_I(x;x') \; . \label{Gravprop}
\end{equation}
We have already seen the de Sitter breaking $A$-type propagator
(\ref{DeltaA}). The $B$-type and $C$-type propagators obey the
equations,
\begin{equation}
\Bigl[ \square \!-\! (D \!-\! 2) H^2\Bigr] i\Delta_B(x;x')
= \frac{i\delta^D(x \!-\! x')}{\sqrt{-g}} = \Bigl[ \square \!-\! 2 (D \!-\! 3)
H^2\Bigr] i\Delta_C(x;x') \; .
\label{bpropeq}
\end{equation}
Each of these propagators is de Sitter invariant and consists of
$i\Delta_{\rm cf}(x;x')$ plus an infinite series of less singular
terms which vanish in $D=4$ dimensions,
\begin{eqnarray}
\lefteqn{i \Delta_B(x;x') = B(y)= i \Delta_{\rm cf}(x;x') - \frac{H^{D-2}}{(4
\pi)^{\frac{D}2}} \! \sum_{n=0}^{\infty}\! \left\{\!  \frac{\Gamma(n \!+\! D
\!-\! 2)}{\Gamma(n \!+\! \frac{D}2)} \Bigl(\frac{y}4 \Bigr)^n \right. }
\nonumber \\
& & \hspace{7.8cm} \left. - \frac{\Gamma(n \!+\!  \frac{D}2)}{\Gamma(n \!+\!
2)} \Bigl( \frac{y}4 \Bigr)^{n - \frac{D}2 +2} \!\right\} \! , \qquad
\label{DeltaB} \\
\lefteqn{i \Delta_C(x;x') = C(y)= i \Delta_{\rm cf}(x;x') +
\frac{H^{D-2}}{(4\pi)^{\frac{D}2}} \! \sum_{n=0}^{\infty} \left\{\!
(n\!+\!1) \frac{\Gamma(n \!+\! D \!-\! 3)}{\Gamma(n \!+\! \frac{D}2)}
\Bigl(\frac{y}4 \Bigr)^n \right. } \nonumber \\
& & \hspace{5cm} \left. - \Bigl(n \!-\! \frac{D}2 \!+\!  3\Bigr) \frac{
\Gamma(n \!+\! \frac{D}2 \!-\! 1)}{\Gamma(n \!+\! 2)} \Bigl(\frac{y}4
\Bigr)^{n - \frac{D}2 +2} \!\right\} \! . \qquad \label{DeltaC}
\end{eqnarray}
Note that the $B$-type and $C$-type propagators agree for $D=4$ dimensions.
The tensor factors are,
\begin{eqnarray}
\Bigl[{}_{\mu\nu} T^A_{\rho\sigma}\Bigr] & = & 2 \, \overline{\eta}_{\mu (\rho}
\overline{\eta}_{\sigma) \nu} - \frac2{D\!-\! 3} \overline{\eta}_{\mu\nu}
\overline{\eta}_{\rho \sigma} \; , \label{TA} \\
\Bigl[{}_{\mu\nu} T^B_{\rho\sigma}\Bigr] & = & -4 \delta^0_{(\mu}
\overline{\eta}_{\nu) (\rho} \delta^0_{\sigma)} \; , \label{TB} \\
\Bigl[{}_{\mu\nu} T^C_{\rho\sigma}\Bigr] & = & \frac2{(D \!-\!2) (D \!-\!3)}
\Bigl[(D \!-\!3) \delta^0_{\mu} \delta^0_{\nu} + \overline{\eta}_{\mu\nu}\Bigr]
\Bigl[(D \!-\!3) \delta^0_{\rho} \delta^0_{\sigma} + \overline{\eta}_{\rho
\sigma}\Bigr] \; , \label{TC}
\end{eqnarray}
where $\overline{\eta}^{\mu\nu} \equiv \eta^{\mu\nu} + \delta^{\mu}_0
\delta^{\nu}_0$ is the spatial part of the Minkowski metric.

The quadratic part of the electromagnetic action is,
\begin{equation}
\mathcal{L}^{(2)}_{\rm EM} = -\frac12 a^{D-4} \partial_{\mu} A_{\nu}
\partial^{\mu} A^{\nu} + \frac12 a^{D-4} \partial_{\mu} A_{\nu}
\partial^{\nu} A^{\mu} \; .
\end{equation}
Of course it is no more possible to add a de Sitter invariant gauge
fixing function for electromagnetism than it is for gravity \cite{MTW1}.
Unlike gravitons, photons show no physical breaking of de Sitter
invariance, so the photon propagator is manifestly de Sitter invariant
if one employs an exact de Sitter invariant gauge condition \cite{TW3}.
However, the de Sitter breaking of the graviton propagator implies that
there is no point to keeping the photon propagator invariant. We have
chosen instead to add the noncovariant gauge fixing function which is
most closely related to the gravitational one (\ref{GRfix}) \cite{RPW},
\begin{equation}
\mathcal{L}_{GF} = -\frac12 a^{D-4} \Bigl(\eta^{\mu\nu} A_{\mu , \nu}
- (D \!-\!4) H a A_0\Bigr)^2 \; . \label{EMfix}
\end{equation}
The associated photon propagator is \cite{RPW},
\begin{equation}
i \Bigl[ \mbox{}_{\mu} \Delta_{\rho}\Bigr](x;x') =
\overline{\eta}_{\mu\rho} \times a a' i\Delta_B(y) - \delta^0_{\mu}
\delta^0_{\rho} \times a a' i\Delta_C(x;x') \; . \label{Photprop}
\end{equation}

\subsection{Our Structure Functions}\label{Struc}

The vacuum polarization is a bi-vector density $i[\mbox{}^{\mu}
\Pi^{\nu}](x;x')$ which is transverse at each point,
\begin{equation}
\frac{\partial}{\partial x^{\mu}} \Bigl[\mbox{}^{\mu}
\Pi^{\nu}\Bigr](x;x') = 0 = \frac{\partial}{\partial {x'}^{\nu}}
\Bigl[\mbox{}^{\mu} \Pi^{\nu}\Bigr](x;x') \; . \label{trans}
\end{equation}
Although it possesses 16 components in $D=4$ dimensions, the
combination of transversality (\ref{trans}), reflection invariance
--- $[\mbox{}^{\mu} \Pi^{\nu}](x;x') = [\mbox{}^{\nu}
\Pi^{\mu}](x';x)$ --- and the coordinate symmetries of the vacuum
relate these components so that they can be expressed in terms of a
very few structure functions. For example, Poincar\'e invariance
implies the following simple form in the flat space limit,
\begin{equation}
\Bigl[\mbox{}^{\mu} \Pi^{\nu}\Bigr]_{\rm flat} = \Bigl(
\eta^{\mu\nu} \eta^{\rho\sigma} \!-\! \eta^{\mu\sigma}
\eta^{\nu\rho} \Bigr) \partial_{\rho} \partial_{\sigma}' \Pi(x \!-\!
x') \; . \label{flatPi}
\end{equation}
The one loop fermion and scalar contributions to $\Pi(x-x')$ have
been known for decades, and an explicit result for the one loop
graviton contribution has recently been derived \cite{LW}.

Because the graviton vacuum on de Sitter background does not respect
full de Sitter invariance, but only spatial homogeneity and
isotropy, it turns out that two structure functions are needed
\cite{LPW1}. We could still choose to represent the transverse
projection operators using covariant derivatives. However, a
detailed examination of this form for the already derived vacuum
polarization from SQED \cite{SQED} reveals that it is cumbersome and
that it obscures rather than simplifies the essential physics
\cite{LPW1}. This seems to be because the conformal invariance of
classical electromagnetism in $D=4$ dimensions is a more powerful
organizing principle that the background's isometries. We have
therefore chosen to employ the noncovariant form originally used to
represent the SQED result,
\begin{equation}
i\Bigl[\mbox{}^{\mu} \Pi^{\nu}\Bigr](x;x') = \Bigl( \eta^{\mu\nu}
\eta^{\rho\sigma} \!-\! \eta^{\mu\sigma} \eta^{\nu\rho} \Bigr)
\partial_{\rho} \partial_{\sigma}' F(x;x') + \Bigl(
\overline{\eta}^{\mu\nu} \overline{\eta}^{\rho\sigma} \!-\!
\overline{\eta}^{\mu\sigma} \overline{\eta}^{\nu\rho} \Bigr)
\partial_{\rho} \partial_{\sigma}' G(x;x') \; . \label{F+G}
\end{equation}
We remind the reader that $\overline{\eta}^{\mu\nu} \equiv
\eta^{\mu\nu} + \delta^{\mu}_0 \delta^{\nu}_0$ is the purely spatial
part of the Minkowski metric.

By comparing (\ref{F+G}) with (\ref{flatPi}) one sees that our
structure function $F(x;x')$ must agree with the flat space result
$i\Pi(x - x')$ in the limit that $H$ vanishes with the co-moving
time $t = \ln(a)/H$ held fixed. Hence the leading divergences are
contained in $F(x;x')$. All terms in $G(x;x')$ must contain at least
one factor of $H^2$, and they are correspondingly less divergent.
Although our representation (\ref{F+G}) is not de Sitter covariant,
the two structure functions have very simple physical
interpretations in terms of changes in the electric permittivity and
the magnetic permeability \cite{AmJ}. However, there is a
straightforward procedure for converting our results for $F(x;x')$
and $G(x;x')$ to the physically opaque, de Sitter covariant
representation \cite{LPW2} if that is desired.

\subsection{Our Counterterms}\label{counter}

Deser and van Nieuwenhuizen showed that Einstein + Maxwell is not
renormalizable at one loop order \cite{SDPvN}. However, it is
straightforward to absorb the divergences, order by order, with BPHZ
counterterms \cite{BPHZ}. We can then solve the quantum-corrected
Maxwell equation (\ref{QMax}) in the standard sense of effective
field theory \cite{Donoghue}. This has already been done for quantum
gravitational corrections to electrodynamics on flat background
\cite{BB,LW}.

The vacuum polarization has two external photon lines so our
computation requires counterterms with two vector potentials. The
superficial degree of divergence is four at one loop order, which
means there must be four derivatives acting either upon vector
potentials or metrics. However, $U(1)$ gauge invariance implies that
at least two of the derivatives must act on vector potentials. If
both of the remaining two derivatives act upon vector potentials we
have the single counterterm which survives in flat space \cite{LW},
\begin{equation}
C_4 D_{\alpha} F_{\mu\nu} D_{\beta} F_{\rho\sigma} g^{\alpha\beta}
g^{\mu\rho} g^{\nu\sigma} \sqrt{-g} \; . \label{flatct}
\end{equation}
Because our gauge fixing functions (\ref{GRfix}) and (\ref{EMfix})
reduce, for $H \rightarrow 0$ at fixed co-moving time, to those
employed in our previous flat space computation \cite{LW}, the
divergent part of $C_4$ must agree as well,
\begin{equation}
C_4 = \frac{\kappa^2}{128 \pi^{\frac{D}2}} \frac{D \Gamma(\frac{D}2
\!-\! 1)}{(D \!-\! 1) (D \!-\! 4)} \; . \label{C4}
\end{equation}

There are three invariant counterterms with two derivatives acting
on two vector potentials and two acting on metrics,
\begin{equation}
C_1 F_{\mu\nu} F_{\rho\sigma} g^{\mu\rho} g^{\nu \sigma} R \sqrt{-g}
\!+\! C_2 F_{\mu\nu} F_{\rho\sigma} g^{\mu\rho} R^{\nu\sigma}
\sqrt{-g} \!+\! C_3 F_{\mu\nu} F_{\rho\sigma} R^{\mu\nu\rho\sigma}
\sqrt{-g} \; .
\end{equation}
However, all the curvatures are related in de Sitter background,
\begin{eqnarray}
R^{\mu\nu\rho\sigma} & \longrightarrow & H^2 a^{-4} \Bigl[
\eta^{\mu\rho} \eta^{\nu\sigma} \!-\! \eta^{\mu\sigma} \eta^{\nu\rho}
\Bigr] \; , \\
R^{\mu\nu} & \longrightarrow & (D \!-\! 1) H^2 a^{-2} \eta^{\mu\nu} \; , \\
R & \longrightarrow & (D \!-\! 1) D H^2 \; .
\end{eqnarray}
Our computation therefore determines only the combination
$\overline{C} \equiv (D-1) D \times C_1 + (D-1) \times C_2 + 2
\times C_3$, and we can write the resulting counterterm as just
$\overline{C} H^2 F_{\mu\nu} F_{\rho\sigma} g^{\mu\rho}
g^{\nu\sigma} \sqrt{-g}$.

Because our gauge fixing functions (\ref{GRfix}) and (\ref{EMfix})
break de Sitter invariance, it is necessary to consider counterterms
which are $U(1)$ invariant but not generally coordinate invariant.
Two properties of our gauge conditions restrict the number of
possibilities:
\begin{itemize}
\item{They become Poincar\'e invariant in the flat
space limit of $H \rightarrow 0$ at fixed co-moving time; and}
\item{They preserve spatial homogeneity and
isotropy, as well as the dilatation symmetry $x^{\mu} \rightarrow k
x^{\mu}$.}
\end{itemize}
The first property means that (\ref{flatct}) is the only counterterm
without explicit factors of $H$. The second property implies that
the only extra counterterm we require has the form, $\Delta C H^2
F_{ij} F_{k\ell} g^{ik} g^{j\ell} \sqrt{-g}$. Our computation
therefore requires only three counterterms,
\begin{eqnarray}
\lefteqn{\mathcal{L}_{\rm BPHZ} = \Delta C H^2 F_{ij} F_{k\ell}
g^{ik} g^{j\ell} \sqrt{-g} \!+\! \overline{C} H^2 F_{\mu\nu}
F_{\rho\sigma} g^{\mu\rho} g^{\nu\sigma} \sqrt{-g} } \nonumber \\
& & \hspace{6cm} + C_4 D_{\alpha} F_{\mu\nu} D_{\beta}
F_{\rho\sigma} g^{\alpha\beta} g^{\mu\rho} g^{\nu\sigma} \sqrt{-g}
\; . \label{LBPHZ} \qquad
\end{eqnarray}

It remains to work out how the counterterms (\ref{LBPHZ}) affect the
structure functions $F(x;x')$ and $G(x;x')$ in our representation
(\ref{F+G}) of the vacuum polarization. The first step is taking the 
variation with respect to $A_{\mu}(x)$ and $A_{\nu}(x')$,
\begin{eqnarray}
\lefteqn{ \frac{\delta S_{\rm BPHZ}}{\delta A_{\mu}(x) \delta
A_{\nu}(x')} = -4 \Delta C H^2 \partial_{\rho} \Biggl\{ \sqrt{-g(x)}
\, \overline{g}^{\mu\gamma}(x) \overline{g}^{\rho\sigma}(x)
\frac{\delta F_{\sigma\gamma}(x)}{\delta A_{\nu}(x')} \Biggr\} }
\nonumber \\
& & \hspace{.5cm} - \partial_{\rho} \Biggl\{ \sqrt{-g(x)} \,
g^{\mu\gamma}(x) g^{\rho\sigma}(x) \Bigl[4 \overline{C} H^2 \!-\! 4
C_4 g^{\alpha\beta}(x) D_{\alpha} D_{\beta} \Bigr] \frac{\delta
F_{\sigma\gamma}(x)}{\delta A_{\nu}(x')} \Biggr\} \; . \qquad 
\end{eqnarray}
The next step is specializing to de Sitter, with $g_{\mu\nu} = a^2 
\eta_{\mu\nu}$ and $\Gamma^{\rho}_{~\mu\nu} = a H (\delta^{\rho}_{\mu}
\delta^0_{\nu} + \delta^{\rho}_{\nu} \delta^0_{\mu} - \eta^{\rho 0}
\eta_{\mu\nu})$. This is very simple for the $\Delta C$ and $\overline{C}$
counterterms,
\begin{eqnarray}
-4 \Delta C H^2 \partial_{\rho} \Biggl\{ \sqrt{-g}
\, \overline{g}^{\mu\gamma} \overline{g}^{\rho\sigma}
\frac{\delta F_{\sigma\gamma}(x)}{\delta A_{\nu}(x')} \Biggr\}
&\!\!\! =\!\!\! & \Bigl( \overline{\eta}^{\mu\nu} \overline{\eta}^{\rho\sigma} 
\!-\! \overline{\eta}^{\mu\sigma} \overline{\eta}^{\rho\nu} \Bigr)
\partial_{\rho} \partial_{\sigma}' \Biggl\{ 4 \Delta C H^2 a^{D-4}
\delta^D(x \!-\! x') \Biggr\} \; , \qquad \label{DG1} \\
-4 \overline{C} H^2 \partial_{\rho} \Biggl\{ \sqrt{-g}
\, g^{\mu\gamma} g^{\rho\sigma} \frac{\delta 
F_{\sigma\gamma}(x)}{\delta A_{\nu}(x')} \Biggr\}
& \!\!\!=\!\!\! & \Bigl( \eta^{\mu\nu} \eta^{\rho\sigma} \!-\!
\eta^{\mu\sigma} \eta^{\rho\nu} \Bigr) \partial_{\rho}
\partial_{\sigma}' \Biggl\{4\overline{C}H^2a^{D-4}\delta^D(x \!-\! x') \Biggr\} \; . \qquad \label{DF1}
\end{eqnarray}
Of course (\ref{DF1}) contributes directly to the structure function
$F(x;x')$, and (\ref{DG1}) contributes to $G(x;x')$. The $C_4$ counterterm
is complicated because of the way the tensor d'Alembertian acts on
$F_{\sigma\gamma}$,
\begin{eqnarray}
\lefteqn{\square F_{\sigma\gamma} = \frac1{a^2} \Biggl\{ \Bigl[ 
\partial^2 \!-\! (D \!-\! 4) H a \partial_0 \!+\! 2 (D-2) H^2 a^2\Bigr] 
F_{\sigma\gamma} \nonumber } \\
& & \hspace{2cm} - 2 H a \Bigl[ \delta^0_{\sigma} \partial^{\alpha} 
F_{\alpha \gamma} \!-\! \delta^0_{\gamma} \partial^{\alpha} F_{\alpha \sigma} 
\Bigr] \!+\! (D \!-\! 4) H^2 a^2 \Bigl[ \delta^0_{\sigma} F_{0 \gamma} \!-\!
\delta^0_{\gamma} F_{0 \sigma} \Bigr] \Biggr\} \; . \qquad \label{DalF}
\end{eqnarray}
The first term of (\ref{DalF}) contributes only to the structure function
$F(x;x')$, whereas the second and third terms contribute to both
structure functions. After some tedious tensor algebra and reflection
of derivatives on the delta function ($\partial'_{\alpha} \delta^D(x - x')
= -\partial_{\alpha} \delta^D(x- x')$), we find that the counterterms
make the following contributions to the two structure functions,
\begin{eqnarray}
\Delta F(x;x') & = & 4 \Bigl[ \overline{C} \!-\! (3D \!-\! 8) C_4\Bigr] 
H^2 a^{D-4} i \delta^D(x \!-\! x') \!-\! 4 C_4 a^{D-6} \Bigl[ 
\partial^2 \!-\! (D\!-\! 6) H a \partial_0 \Bigr] \, i \delta^D(x \!-\! x') 
\; , \qquad \label{DeltaF} \\
\Delta G(x;x') & = & 4 \Bigl[\Delta C \!-\! (D \!-\!6) C_4 \Bigr] 
H^2 a^{D-4} i \delta^D(x \!-\! x') \; . \qquad \label{DeltaG}
\end{eqnarray}
The simplicity of this form is one more indication, added to the 
many already found \cite{LPW1}, of the superiority of the noncovariant
representation over the covariant one.

%---------------------------------------------------------------------------------------------------------------------------------------------------------------%

\section{The 4-Point Contribution}\label{4point}

This section summarizes the derivation of the contribution to the vacuum polarization from the 4-point diagram in Fig.~\ref{photon}, by deriving the structure functions required for the desired representation of the vacuum polarization (\ref{F+G}). This is achieved by first completing all naive index contractions of (\ref{4pt1}), followed by a substitution of the graviton propagator, which allows all final index contractions to be taken. Lastly, several contraction identities are introduced from which the 4-point structure functions can be deduced. The procedure outlined in this section will be similar to that used for deriving the 3-point contribution and serves as a simple guide to extracting the desired structure functions. 

\subsection{Naive Contractions}

In order to derive the two scalar structure functions from the 4-point contribution to the vacuum polarization (\ref{4pt1}) the first step will be to substitute (\ref{Uvert}) for the 4-point vertex function and apply the naive contractions. This will naturally engender many terms, and to make the process simple for the reader to follow we will break (\ref{Uvert}) into pieces and then define the 4-point contribution as a sum of those pieces
\be
i\Bigl[\mbox{}^{\mu} \Pi^{\nu}_{\rm 4pt}\Bigr](x;x') =
\partial_{\kappa} \Biggl\{ i\kappa^2 a^{D-4}
\sum\limits_{I=1}^{6} U_I^{\mu\nu\kappa\lambda\alpha\beta\gamma\delta}
\, i\Bigl[\mbox{}_{\alpha\beta} \Delta_{\gamma\delta} \Bigr](x;x)
\, \partial_{\lambda} \delta^D(x \!-\! x') \Biggr\} \; , \label{4pt2}
\ee
where Table \ref{4parts} lists the various $U_I^{\mu\nu\kappa\lambda\alpha\beta\gamma\delta}$.
\begin{table}
\centering
\begin{tabular}{|c|c|}
\hline
I & $U_I^{\mu\rho\kappa\lambda\alpha\beta\gamma\delta}$  \Tstrut\Bstrut\\
\hline
1 & $\frac{1}{4}\e^{\k[\lam}\e^{\rho]\mu}\e^{\a\b}\e^{\g\d}$ \Tstrut\Bstrut\\\
2 & $-\frac{1}{2}\e^{\k[\lam}\e^{\rho]\mu}\e^{\a(\g}\e^{\d)\b}$ \Tstrut\Bstrut\\\
3 & $\e^{\a\b}\e^{\g)[\mu}\e^{\k][\rho}\e^{\lam](\d}+\e^{\g\d}\e^{\a)[\mu}\e^{\k][\rho}\e^{\l](\b}$ \Tstrut\Bstrut\\\
4 & $\e^{\k(\a}\e^{\b)[\l}\e^{\rho](\g}\e^{\d)\mu}+\eta^{\kappa (\gamma}
\eta^{\delta) [\lambda} \eta^{\rho ] (\alpha} \eta^{\beta) \mu}$ \Tstrut\Bstrut\\\
5 & $\eta^{\kappa (\alpha} \eta^{\beta) (\gamma} \eta^{\delta) [\lambda}
\eta^{\rho ] \mu}+\eta^{\kappa (\gamma} \eta^{\delta) (\alpha}
\eta^{\beta) [\lambda} \eta^{\rho ] \mu}$ \Tstrut\Bstrut\\\
6 & $\eta^{\kappa [\lambda}
\eta^{\rho ] (\alpha} \eta^{\beta) (\gamma} \eta^{\delta) \mu}+\eta^{\kappa [\lambda} \eta^{\rho ] (\gamma} \eta^{\delta) (\alpha}
\eta^{\beta) \mu}$ \Tstrut\Bstrut\\
\hline
\end{tabular}
\caption{Parts of 4-point vertex function \label{4parts}}
\end{table}
Once the vertex function terms have been inserted into the vacuum polarization the naive index contractions can be carried out term by term, the results of which are listed in Table \ref{4naive}.
\begin{table}
\centering
\begin{tabular}{|c|l|}
\hline 
I & $\partial_{\kappa} \Biggl\{ i\kappa^2 a^{D-4}
U_I^{\mu\nu\kappa\lambda\alpha\beta\gamma\delta}
\, i\Bigl[\mbox{}_{\alpha\beta} \Delta_{\gamma\delta} \Bigr](x;x)
\, \partial_{\lambda} \delta^D(x \!-\! x') \Biggr\}$  \Tstrut\Bstrut\\
\hline
1 & $\frac{1}{8}\e^{\mu\nu}\del_\k\left\{i\k^2a^{D-4}i\Bigl[\mbox{}^\b\;_\b\Delta^\d\;_\d \Bigr]\del^\k\d^D\right\}-\frac{1}{8}\del^\nu\left\{i\k^2a^{D-4}i\Bigl[\mbox{}^\b\;_\b\Delta^\d\;_\d \Bigr]\del^\mu\d^D\right\}$ \Tstrut\Bstrut\\
\hline 
2 & $-\frac{1}{4}\e^{\mu\nu}\del_\k\left\{i\k^2a^{D-4}i\Bigl[\mbox{}^{\g\d}\Delta_{\g\d} \Bigr]\del^\k\d^D\right\}+\frac{1}{4}\del^\nu\left\{i\k^2a^{D-4}i\Bigl[\mbox{}^{\g\d}\Delta_{\g\d} \Bigr]\del^\mu\d^D\right\}$ \Tstrut\Bstrut\\
\hline
3 & $-\frac{1}{2}\e^{\mu\nu}\del_\k\left\{i\k^2a^{D-4}i\Bigl[\mbox{}^\b\;_\b\Delta^{\k\l} \Bigr]\del_\l\d^D\right\}+\frac{1}{2}\del^\nu\left\{i\k^2a^{D-4}i\Bigl[\mbox{}^\b\;_\b\Delta^{\mu\l} \Bigr]\del_\l\d^D\right\}$ \Tstrut\Bstrut\\
 & $-\frac{1}{2}\del_\k\left\{i\k^2a^{D-4}i\Bigl[\mbox{}^\b\;_\b\Delta^{\mu\nu} \Bigr]\del^\k\d^D\right\}+\frac{1}{2}\del_\k\left\{i\k^2a^{D-4}i\Bigl[\mbox{}^\b\;_\b\Delta^{\k\nu} \Bigr]\del^\mu\d^D\right\}$ \Tstrut\Bstrut\\
\hline
4 & $\del_\k\left\{i\k^2a^{D-4}i\Bigl[\mbox{}^{\mu\nu}\Delta^{\k\l} \Bigr]\del_\l\d^D\right\}-\del_\k\left\{i\k^2a^{D-4}i\Bigl[\mbox{}^{\l\mu}\Delta^{\k\nu} \Bigr]\del_\l\d^D\right\}$ \Tstrut\Bstrut\\
\hline
5 & $\e^{\mu\nu}\del_\k\left\{i\k^2a^{D-4}i\Bigl[\mbox{}^{\l\d}\Delta^\k\;_\d \Bigr]\del_\l\d^D\right\}-\del_\k\left\{i\k^2a^{D-4}i\Bigl[\mbox{}^{\nu\d}\Delta^\k\;_\d \Bigr]\del^\mu\d^D\right\}$ \Tstrut\Bstrut\\
\hline
6 & $\del_\k\left\{i\k^2a^{D-4}i\Bigl[\mbox{}^{\mu\d}\Delta^\nu\;_\d \Bigr]\del^\k\d^D\right\}-\del^\nu\left\{i\k^2a^{D-4}i\Bigl[\mbox{}^{\mu\d}\Delta^\l\;_\d \Bigr]\del_\l\d^D\right\}$ \Tstrut\Bstrut\\
\hline
\end{tabular}
\caption{Terms of 4-point contribution after naive index contractions. \label{4naive}}
\end{table}

\subsection{Substitution of Graviton Propagator}\label{subgravprop}

To complete the index contractions the full graviton propagator must be inserted. This will again create many more terms so it is useful to break the graviton propagator into three pieces and consider each part separately. Upon consideration of the graviton propagator (\ref{Gravprop}) we see that if each of the three types of scalar propagators are set equal, to say $B(y)$, then the tensor components combine to give the conformal graviton propagator tensor component
\be
\Bigl[\mbox{}_{\mu\nu}T^{\rm A}_{\rho\si}\Bigr]+\Bigl[\mbox{}_{\mu\nu}T^{\rm B}_{\rho\si}\Bigr]+\Bigl[\mbox{}_{\mu\nu}T^{\rm C}_{\rho\si}\Bigr]=\Bigl[\mbox{}_{\mu\nu}T^{\rm cf}_{\rho\si}\Bigr]=2\e_{\mu(\rho}\e_{\si)\nu}-\frac{2}{(D-2)}\e_{\mu\nu}\e_{\rho\si} \; .
\ee
By adding and subtracting $B(y)$ from each of the scalar propagators in (\ref{Gravprop}) the graviton propagator can be rewritten as
\bea
i\Bigl[\mbox{}_{\mu\nu}\D_{\rho\si}\Bigr](x;x')&=&\Bigl[\mbox{}_{\mu\nu}T^{\rm cf}_{\rho\si}\Bigr]\times B+\Bigl[\mbox{}_{\mu\nu}T^{A}_{\rho\si}\Bigr]\times \left(i\D_A-B\right)+\Bigl[\mbox{}_{\mu\nu}T^{\rm C}_{\rho\si}\Bigr]\times \left(C-B\right) \; .
\label{Gravprop2}
\eea

Writing the graviton propagator in this way cancels the conformal parts of the scalar propagators in the second two terms. This property is not useful for the 4-point contribution, but it will be helpful for renormalizing the 3-point contribution and we use it again here for consistency. We will refer to the three components of (\ref{Gravprop2}) separately as the conformal part, the A-type part and and C-type part as associated with the tensor factor of each piece. 

Before making substitutions for the graviton propagator we need to know how each of the tensor components contract in the various ways appearing in Table \ref{4naive}. The relevant contractions are listed in Table \ref{gravcont}, where each element represents the tensor factor in the top row contracted with the combination of flat space metrics listed in the left hand column.
\begin{table}
\centering
\begin{tabular}{|c|ccc|}
\hline
 & $\Bigl[\mbox{}_{\mu\nu}T^{\rm cf}_{\rho\si}\Bigr]$ & $\Bigl[\mbox{}_{\mu\nu}T^{A}_{\rho\si}\Bigr]$ & $\Bigl[\mbox{}_{\mu\nu}T^{\rm C}_{\rho\si}\Bigr]$ \Tstrut\Bstrut\\
\hline
$\e^{\mu\nu}\e^{\rho\si}$ & $-\frac{4D}{(D-2)}$ & $-\frac{4(D-1)}{(D-3)}$ & $\frac{8}{(D-2)(D-3)}$ \Tstrut\Bstrut\\
$\e^{\mu\rho}\e^{\nu\si}$ & $\frac{D(D^2-D-4)}{(D-2)}$ & $\frac{(D^3-4D^2+D+2)}{(D-3)}$ & $2\frac{(D^2-5D+8)}{(D-2)(D-3)}$ \Tstrut\Bstrut\\
$\e^{\mu\nu}$ & $-\frac{4}{(D-2)}\e_{\rho\si}$ & $-\frac{4}{(D-3)}\bar{\e}_{\rho\si}$ & $\frac{4}{(D-2)}\d^0_\rho\d^0_\si+\f{4}{(D-2)(D-3)}\bar{\e}_{\rho\si}$ \Tstrut\Bstrut\\
$\e^{\mu\rho}$ & $\frac{(D^2-D-4)}{(D-2)}\e_{\nu\si}$ & $\frac{(D^2-3D-2)}{(D-3)}\bar{\e}_{\nu\si}$ & $-2\frac{(D-3)}{(D-2)}\d^0_\nu\d^0_\si+\frac{2}{(D-2)(D-3)}\bar{\e}_{\nu\si}$ \Tstrut\Bstrut\\
\hline
\end{tabular}
\caption{Various graviton tensor factor contractions. \label{gravcont}}
\end{table}

As stated above we dissect the graviton propagator substitution into three parts. The substitution and following index contractions for the conformal part of the graviton propagator are listed in Table \ref{subcf}, the terms resulting from the A-type part are listed in Table \ref{subA}, and the terms from the C-type part in Table \ref{subC}. Note that in these tables we suppress the $(x-x')$ factor on the delta functions for brevity.
\begin{table}
\centering
\begin{tabular}{|c|l|}
\hline 
I & $\partial_{\kappa} \Biggl\{ i\kappa^2 a^{D-4}
U_I^{\mu\nu\kappa\lambda\alpha\beta\gamma\delta}
\, \Bigl[\mbox{}_{\alpha\beta} T^{\rm cf}_{\gamma\delta} \Bigr]B(0)
\, \partial_{\lambda} \delta^D(x \!-\! x') \Biggr\}$  \Tstrut\Bstrut\\
\hline
1 & $-\frac{D}{2(D-2)}\left\{\e^{\mu\nu}\del_\k\left[i\k^2a^{D-4}B\del^\k\d^D\right]-\del^\nu\left[i\k^2a^{D-4}B\del^\mu\d^D\right]\right\}$ \Tstrut\Bstrut\\
\hline 
2 & $-\frac{D(D^2-D-4)}{4(D-2)}\left\{\e^{\mu\nu}\del_\k\left[i\k^2a^{D-4}B\del^\k\d^D\right]-\del^\nu\left[i\k^2a^{D-4}B\del^\mu\d^D\right]\right\}$ \Tstrut\Bstrut\\
\hline
3 & $\frac{4}{(D-2)}\left\{\e^{\mu\nu}\del_\k\left[i\k^2a^{D-4}B\del^\k\d^D\right]-\del^\nu\left[i\k^2a^{D-4}B\del^\mu\d^D\right]\right\}$ \Tstrut\Bstrut\\
\hline
4 & $-\frac{D}{(D-2)}\left\{\e^{\mu\nu}\del_\k\left[i\k^2a^{D-4}B\del^\k\d^D\right]-\del^\nu\left[i\k^2a^{D-4}B\del^\mu\d^D\right]\right\}$ \Tstrut\Bstrut\\
\hline
5 & $\frac{(D^2-D-4)}{(D-2)}\left\{\e^{\mu\nu}\del_\k\left[i\k^2a^{D-4}B\del^\k\d^D\right]-\del^\nu\left[i\k^2a^{D-4}B\del^\mu\d^D\right]\right\}$ \Tstrut\Bstrut\\
\hline
6 & $\frac{(D^2-D-4)}{(D-2)}\left\{\e^{\mu\nu}\del_\k\left[i\k^2a^{D-4}B\del^\k\d^D\right]-\del^\nu\left[i\k^2a^{D-4}B\del^\mu\d^D\right]\right\}$ \Tstrut\Bstrut\\
\hline
\end{tabular}
\caption{4-point contributions coming from the conformal part of the graviton propagator. \label{subcf}}
\end{table}
Upon further inspection of the conformal part of the graviton propagator in Table \ref{subcf} we see that all six sets of terms have the same tensor structure. Combining all of the terms we find the simplified expression
\bea
i\Bigl[\mbox{}^{\mu} \Pi^{\nu}_{\rm 4pt}\Bigr]_{\rm cf}(x;x') &=&
-\frac{(D^3-9D^2+10D+16)}{4(D-2)}\left\{\e^{\mu\nu}\del_\k\left[i\k^2a^{D-4}B(0)\del^\k\d^D(x-x')\right]\right. \nonumber \\
&&\left.-\del^\nu\left[i\k^2a^{D-4}B(0)\del^\mu\d^D(x-x')\right]\right\} \; .\label{4ptcf}
\eea
\begin{table}
\centering
\begin{tabular}{|c|l|}
\hline 
I & $\partial_{\kappa} \Biggl\{ i\kappa^2 a^{D-4}
U_I^{\mu\nu\kappa\lambda\alpha\beta\gamma\delta}
\, \Bigl[\mbox{}_{\alpha\beta} T^{\rm A}_{\gamma\delta} \Bigr]\left(i\D_A(x;x)-B(0)\right)
\, \partial_{\lambda} \delta^D(x \!-\! x') \Biggr\}$  \Tstrut\Bstrut\\
\hline
1 & $-\frac{(D-1)}{2(D-3)}\left\{\e^{\mu\nu}\del_\k\left[i\k^2a^{D-4}(i\D_A-B)\del^\k\d^D\right]-\del^\nu\left[i\k^2a^{D-4}(i\D_A-B)\del^\mu\d^D\right]\right\}$ \Tstrut\Bstrut\\
\hline 
2 & $-\frac{(D^3-4D^2+D+2)}{4(D-3)}\left\{\e^{\mu\nu}\del_\k\left[i\k^2a^{D-4}(i\D_A-B)\del^\k\d^D\right]-\del^\nu\left[i\k^2a^{D-4}(i\D_A-B)\del^\mu\d^D\right]\right\}$ \Tstrut\Bstrut\\
\hline
3 & $\frac{2}{(D-3)}\left\{\e^{\mu\nu}\bar{\del}_\k\left[i\k^2a^{D-4}(i\D_A-B)\bar{\del}^\k\d^D\right]-\del^\nu\left[i\k^2a^{D-4}(i\D_A-B)\bar{\del}^\mu\d^D\right]\right.$ \Tstrut\Bstrut\\
 & $\quad\quad\quad\left.+\bar{\e}^{\mu\nu}\del_\k\left[i\k^2a^{D-4}(i\D_A-B)\del^\k\d^D\right]-\bar{\del}^\nu\left[i\k^2a^{D-4}(i\D_A-B)\del^\mu\d^D\right]\right\}$ \Tstrut\Bstrut\\
\hline
4 & $-\frac{(D-1)}{(D-3)}\left\{\bar{\e}^{\mu\nu}\bar{\del}_\k\left[i\k^2a^{D-4}(i\D_A-B)\bar{\del}^\k\d^D\right]-\bar{\del}^\nu\left[i\k^2a^{D-4}(i\D_A-B)\bar{\del}^\mu\d^D\right]\right\}$ \Tstrut\Bstrut\\
\hline
5 & $\frac{(D^2-3D-2)}{(D-3)}\left\{\e^{\mu\nu}\bar{\del}_\k\left[i\k^2a^{D-4}(i\D_A-B)\bar{\del}^\k\d^D\right]-\bar{\del}^\nu\left[i\k^2a^{D-4}(i\D_A-B)\del^\mu\d^D\right]\right\}$ \Tstrut\Bstrut\\
\hline
6 & $\frac{(D^2-3D-2)}{(D-3)}\left\{\bar{\e}^{\mu\nu}\del_\k\left[i\k^2a^{D-4}(i\D_A-B)\del^\k\d^D\right]-\del^\nu\left[i\k^2a^{D-4}(i\D_A-B)\bar{\del}^\mu\d^D\right]\right\}$ \Tstrut\Bstrut\\
\hline
\end{tabular}
\caption{4-point contributions coming from the A-type part of the graviton propagator. \label{subA}}
\end{table}
\begin{table}
\centering
\begin{tabular}{|c|l|}
\hline 
I & $\partial_{\kappa} \Biggl\{ i\kappa^2 a^{D-4}
U_I^{\mu\nu\kappa\lambda\alpha\beta\gamma\delta}
\, \Bigl[\mbox{}_{\alpha\beta} T^{\rm C}_{\gamma\delta} \Bigr](C(0)-B(0))
\, \partial_{\lambda} \delta^D(x \!-\! x') \Biggr\}$  \Tstrut\Bstrut\\
\hline
1 & $\frac{1}{(D-2)(D-3)}\left\{\e^{\mu\nu}\del_\k\left[i\k^2a^{D-4}(C-B)\del^\k\d^D\right]-\del^\nu\left[i\k^2a^{D-4}(C-B)\del^\mu\d^D\right]\right\}$ \Tstrut\Bstrut\\
\hline 
2 & $-\frac{(D^2-5D+8)}{2(D-2)(D-3)}\left\{\e^{\mu\nu}\del_\k\left[i\k^2a^{D-4}(C-B)\del^\k\d^D\right]-\del^\nu\left[i\k^2a^{D-4}(C-B)\del^\mu\d^D\right]\right\}$ \Tstrut\Bstrut\\
\hline
3 & $\frac{2}{(D-2)}\left\{\e^{\mu 0}\del^\nu\left[i\k^2a^{D-4}(C-B)\del^0\d^D\right]+\frac{1}{(D-3)}\del^\nu\left[i\k^2a^{D-4}(C-B)\bar{\del}^\mu\d^D\right]\right.$ \Tstrut\Bstrut\\
 & $-\e^{\mu\nu}\del^0\left[i\k^2a^{D-4}(C-B)\del^0\d^D\right]-\frac{1}{(D-3)}\e^{\mu\nu}\bar{\del}_\k\left[i\k^2a^{D-4}(C-B)\bar{\del}^\k\d^D\right]$ \Tstrut\Bstrut\\
 & $+\e^{\nu 0}\del^0\left[i\k^2a^{D-4}(C-B)\del^\mu\d^D\right]+\frac{1}{(D-3)}\bar{\del}^\nu\left[i\k^2a^{D-4}(C-B)\del^\mu\d^D\right]$ \Tstrut\Bstrut\\
 & $\left.-\e^{\mu 0}\e^{\nu 0}\del_\k\left[i\k^2a^{D-4}(C-B)\del^\k\d^D\right]-\frac{1}{(D-3)}\bar{\e}^{\mu\nu}\del_\k\left[i\k^2a^{D-4}(C-B)\del^\k\d^D\right]\right\}$ \Tstrut\Bstrut\\
\hline
4 & $\frac{2}{(D-2)(d-3)}\left\{(D-3)\bar{\e}^{\mu\nu}\del^0\left[i\k^2a^{D-4}(C-B)\del^0\d^D\right]+\bar{\e}^{\mu\nu}\bar{\del}_\k\left[i\k^2a^{D-4}(C-B)\bar{\del}^\k\d^D\right]\right.$ \Tstrut\Bstrut\\
 & $+(D-3)\e^{\mu 0}\e^{\nu 0}\bar{\del}_\k\left[i\k^2a^{D-4}(C-B)\bar{\del}^\k\d^D\right]-(D-3)\e^{\nu 0}\del^0\left[i\k^2a^{D-4}(C-B)\bar{\del}^\mu\d^D\right]$ \Tstrut\Bstrut\\
 & $\left.-(D-3)\e^{\mu 0}\bar{\del}^\nu\left[i\k^2a^{D-4}(C-B)\del^0\d^D\right]-\bar{\del}^\nu\left[i\k^2a^{D-4}(C-B)\bar{\del}^\mu\d^D\right]\right\}$ \Tstrut\Bstrut\\
\hline
5 & $\frac{2}{(D-2)(D-3)}\left\{-(D-3)^2\e^{\mu\nu}\del^0\left[i\k^2a^{D-4}(C-B)\del^0\d^D\right]-\bar{\del}^\nu\left[i\k^2a^{D-4}(C-B)\del^\mu\d^D\right]\right.$ \Tstrut\Bstrut\\
 & $\left.+(D-3)^2\e^{\nu 0}\del^0\left[i\k^2a^{D-4}(C-B)\del^\mu\d^D\right]+\e^{\mu\nu}\bar{\del}_\k\left[i\k^2a^{D-4}(C-B)\bar{\del}^\k\d^D\right]\right\}$ \Tstrut\Bstrut\\
\hline
6 & $\frac{2}{(D-2)(D-3)}\left\{\bar{\e}^{\mu\nu}\del_\k\left[i\k^2a^{D-4}(C-B)\del^\k\d^D\right]-\del^\nu\left[i\k^2a^{D-4}(C-B)\bar{\del}^\mu\d^D\right]\right.$ \Tstrut\Bstrut\\
 & $\left.-(D-3)^2\e^{\mu 0}\e^{\nu 0}\del_\k\left[i\k^2a^{D-4}(C-B)\del^\k\d^D\right]+(D-3)^2\e^{\mu 0}\del^\nu\left[i\k^2a^{D-4}(C-B)\del^0\d^D\right]\right\}$ \Tstrut\Bstrut\\
\hline
\end{tabular}
\caption{4-point contributions coming from the C-type part of the graviton propagator. \label{subC}}
\end{table}

Equation (\ref{4ptcf}), and Tables \ref{subA} and \ref{subC} make up the 4-point contribution to the vacuum polarization. Next we will transform these results into the desired manifestly transverse form.

\subsection{Finding the 4-Point Structure Functions}\label{4sf}

Recall that it is our goal to write the result for the vacuum polarization in the form of (\ref{F+G}), where two transverse projection operators are acting on two scalar structure functions $F(x;x')$ and $G(x;x')$. Using the Tables of section \ref{subgravprop} we can now derive these structure functions.

The easiest way we found to extract the structure functions was to exploit the known transversality of the vacuum polarization and isolate the structure functions via two contractions, resulting in two equations for the two structure functions. The first contraction is made with $\d_\mu^0\d^0_\nu$, applied to equation (\ref{F+G}) this gives the identity
\be
i\Bigl[\mbox{}^0\Pi^0\Bigl](x;x')=\nabla^2F(x;x') \; ,
\label{00id}
\ee
which provides a simple equation for $F$. To find $G$ the second contraction is taken with $\d_\mu^i\d_\nu^j(j\neq i)$, applied to (\ref{F+G}) we find the second identity
\be
i\Bigl[\mbox{}^i\Pi^j\Bigl](x;x')=-\del'^i\del^j\left[F(x;x')+G(x;x')\right] \; .
\label{ijid}
\ee
Knowing $F$ it is now trivial to find $G$. 

It is true that there are other pairs of contractions that would work equally as well. For example, contracting with $\d^0_\mu\d^i_\nu$ and $\bar{\e}_{\mu\nu}$ leads to the identities
\bea
i\Bigl[\mbox{}^0\Pi^i\Bigl](x;x')&=&-\del'^0\del^iF(x;x') \; , \\
i\Bigl[\mbox{}_{\bar{\mu}}\Pi^{\bar{\mu}}\Bigl](x;x')&=&-(D-1)\del^0\del'^0F(x;x')-(D-2)\nabla^2\left[F(x;x')+G(x;x')\right] \; .
\eea
These identities provide a useful check on the structure functions derived from identities (\ref{00id}) and (\ref{ijid}). There are likely more contractions that would provide similar sets of equations, but these two sets were sufficient for our purposes. 

It is quite a tedious task to show how applying these contractions plays out for the entire 4-point contribution, and the procedure is extremely repetitive. For the reader's sanity and our own we will work out one example and assume that the procedure for the rest of the 4-point contribution can be easily deduced. We will demonstrate how to find $F$ and $G$ from the conformal part of the graviton propagator as given in (\ref{4ptcf}).

Contracting $\d_\mu^0\d_\nu^0$ with (\ref{4ptcf}) we have
\bea
i\Bigl[\mbox{}^0\Pi_{\rm 4pt}^0\Bigl]_{\rm cf}(x;x')&=&-\frac{(D^3-9D^2+10D+16)}{4(D-2)}\Bigl\{-\del_\k\left[i\k^2a^{D-4}B\del^\k\d^D(x-x')\right] \nonumber \\
&&\hspace{2.7in}-\del^0\left[i\k^2a^{D-4}B\del^0\d^D(x-x')\right]\Bigr\} \nonumber \\
&=&-\frac{(D^3-9D^2+10D+16)}{4(D-2)}\Bigl\{\left(\del\cdot\del'+\del^0\del'^0\right)\times\left[i\k^2a^{D-4}B\d^D(x-x')\right]\Bigr\} \nonumber \\
&=&\frac{(D^3-9D^2+10D+16)}{4(D-2)}\nabla^2\left[i\k^2a^{D-4}B(0)\d^D(x-x')\right] \; ,
\eea
where in going from the first line to the second we used the delta function to make the change $\del\rightarrow-\del'$. It is then trivial to pull the inner derivative outside the curly brackets since all prefactors are only functions of $x$. Also the B-type propagator can be evaluated at $y=0$ since it is being evaluated at coincidence $(x=x')$. Upon comparison with identity (\ref{00id}) we find the first structure function
\be
F_{4,{\rm cf}}(x;x')=\f{(D^3-9D^2+10D+16)}{4(D-2)}\left[i\k^2a^{D-4}B(0)\d^D(x-x')\right] \;.
\ee
To find the second structure function we contract $\d_\mu^i\d_\nu^j(j\neq i)$ with (\ref{4ptcf})
\bea
i\Bigl[\mbox{}^i\Pi^j\Bigl]_{\rm cf}(x;x')&=&-\frac{(D^3-9D^2+10D+16)}{4(D-2)}\Bigl\{0-\del^j\left[i\k^2a^{D-4}B\del^i\d^D(x-x')\right]\Bigr\} \nonumber \\
&=&-\frac{(D^3-9D^2+10D+16)}{4(D-2)}\del'^i\del^j\left[i\k^2a^{D-4}B\d^D(x-x')\right] \; .
\eea
Invoking identity (\ref{ijid}) and making the proper substitution for $F_{4,{\rm cf}}$ it is easy to see
\be
G_{4,{\rm cf}}(x;x')=0 \; .
\ee

This concludes the example case for finding $F$ and $G$. The same procedure can be applied to all of the terms in Tables \ref{subA} and \ref{subC} to find the rest of the structure functions. Combining all $F$ and $G$ contributions from the three parts of the graviton propagator produces the full result for the 4-point structure functions
\bea
F_{\rm 4}(x;x')&=&\k^2a^{D-4}\Bigl\{\f{D(D\!-\!5)}{4}i\D_A(0)-\f{(D\!-\!1)}{2}B(0)-\f{(3D\!-\!10)}{2(D\!-\!2)}C(0)\Bigr\}i\d^D(x\!-\!x') \label{f4} \\
&& \nonumber \\
G_{\rm 4}(x;x')&=&-\k^2a^{D-4}\Bigl\{\f{(D^2\!-\!4D\!+\!1)}{(D\!-\!3)}i\D_A(0)-(D\!-\!3)B(0)-2\f{(D\!-\!4)}{(D\!-\!3)}C(0)\Bigr\}i\d^D(x\!-\!x') \label{g4}
\eea 
We can now make substitutions for the propagators. The B- and C-type propagators are a finite constant at coincidence in $D=4$, but the A-type propagator is divergent. Therefore, it is useful to break the structure functions into their finite and divergent pieces
\bea
F_{4,\rm div}(x;x')&=&-\f{D(D-5)}{4}\f{\k^2a^{D-4}H^{D-2}\G(D-1)}{(4\pi)^{D/2}\G(\f{D}{2})}\pi\cot\left(\f{\pi}{2}D\right)i\d^D(x-x') \; , \label{f4div} \\
&& \nonumber \\
F_{4,\rm finite}(x;x')&=&\f{\k^2H^2}{4\pi^2}\left\{\f{1}{4}-\ln(a)\right\}i\d^4(x-x') \; , \label{f4fin} \\
&& \nonumber \\
G_{4,\rm div}(x;x')&=&\f{(D^2-4D+1)}{(D-3)}\f{\k^2a^{D-4}H^{D-2}\G(D-1)}{(4\pi)^{D/2}\G(\f{D}{2})}\pi\cot\left(\f{\pi}{2}D\right)i\d^D(x-x') \; , \label{g4div} \\
&& \nonumber \\
G_{4,\rm finite}(x;x')&=&-\f{\k^2H^2}{4\pi^2}\left\{\f{1}{4}+\ln(a)\right\}i\d^4(x-x') \; . \label{g4fin}
\eea
This concludes our derivation of the 4-point contributions to the structure functions, where we note that only the terms proportional to $\ln(a)$ in (\ref{f4div}-\ref{g4fin}) cannot be absorbed into the counter terms. We will now derive the 3-point contribution.

\section{The 3-Point Contribution}\label{3point}

This section will cover the derivation of the contribution to the vacuum polarization attributed to the 3-point diagram in Fig.~\ref{photon}. The procedure for deriving the scalar structure functions from the diagram is very similar to the one used in section \ref{4point}. First the naive index contractions are completed in pieces by dividing the 3-point vertex function appropriately. Then once the proper substitutions have been made for the graviton and photon propagators all remaining index contractions can be completed. Finally the 3-point contributions to $F$ and $G$ will be presented in several parts, broken up according to which pieces of the photon and graviton propagators the structure function originated from.

\subsection{Naive contractions}

Following the same organizing procedure as in the previous section, we complete the naive index contractions first by breaking the 3-point vertex function (\ref{Vvert}) into pieces, shown in Table \ref{3parts}, and rewriting the 3-point contribution as a sum of these terms
\bea
i\Bigl[\mbox{}^\mu\Pi^\nu_{\rm 3pt}\Bigr](x;x')&=&\del_\k\del'_\th\Bigl\{i\k a^{D-4}\sum\limits_{I=1}^2V_I^{\mu\rho\k\l\a\b}i\Bigl[\mbox{}_{\a\b}\D_{\g\d}\Bigr](x;x') \nonumber \\
&&\hspace{1in}\times i\k a'^{D-4}\sum\limits_{J=1}^2V_J^{\nu\si\th\phi\g\d}\del_\l\del'_\phi i\Bigl[\mbox{}_\rho\D_\si\Bigr](x;x')\Bigl\} \; . \label{3ptsum}
\eea
\begin{table}
\centering
\begin{tabular}{|c|c|}
\hline
I & $V^{\mu\rho\k\l\a\b}$ \Tstrut\Bstrut\\
\hline
1 & $\e^{\a\b}\e^{\k[\l}\e^{\rho]\mu}$ \Tstrut\Bstrut\\
2 & $4\e^{\a)[\mu}\e^{\k][\rho}\e^{\l](\b}$ \Tstrut\Bstrut\\
\hline 
\end{tabular}
\caption{Pieces of 3-point vertex function.  \label{3parts}}
\end{table}
(\ref{3ptsum}) contains products of the parts of the two 3-point vertex functions; the result for completing the naive index contractions for all said products are shown in Table \ref{3naive}. To simplify the results in this table a short hand notation for antisymmetrization has been adopted, where both square brackets and double square brackets imply antisymmetrization. The index structure for these terms can be confusing and it should be noted that the antisymmetrization only applies to the immediate indices with in the brackets. Here is a worked out example
\be
i\Bigl[\mbox{}^{\rho[\mu}\D^{\si[[\nu}\Bigl]\del^{\k]}\del'^{\th]]}=\f{1}{4}\left\{i\Bigl[\mbox{}^{\rho\mu}\D^{\si\nu}\Bigl]\del^{\k}\del'^{\th}-i\Bigl[\mbox{}^{\rho\k}\D^{\si\nu}\Bigl]\del^{\mu}\del'^{\th}+i\Bigl[\mbox{}^{\rho\k}\D^{\si\th}\Bigl]\del^{\mu}\del'^{\nu}-i\Bigl[\mbox{}^{\rho\mu}\D^{\si\th}\Bigl]\del^{\k}\del'^{\nu}\right\} \; .
\ee
\begin{table}
\centering
\begin{tabular}{|c|c|c|}
\hline
I & J & $-\del_\k\del'_\th\Bigl\{\k^2 (aa')^{D-4}V_I^{\mu\rho\k\l\a\b}i\Bigl[\mbox{}_{\a\b}\D_{\g\d}\Bigr](x;x')V_J^{\nu\si\th\phi\g\d}\del_\l\del'_\phi i\Bigl[\mbox{}_\rho\D_\si\Bigr](x;x')\Bigl\}$ \Tstrut\Bstrut\\
\hline
1 & 1 & $-\k^2\del_\k\del'_\th\Bigl\{(aa')^{D-4}i\Bigl[\mbox{}^\b\;_\b\D^\d\;_\d\Bigl]\del^{[\mu}\del'^{[[\nu}i\Bigl[\mbox{}^{\k]}\D^{\th]]}\Bigl]\Bigl\}$ \Tstrut\Bstrut\\
\hline
1 & 2 & $2\k^2\del_\k\del'_\th\Bigl\{(aa')^{D-4}i\Bigl[\mbox{}^\b\;_\b\D^{\phi[\nu}\Bigl]\del^{[[\mu}\del'_\phi i\Bigl[\mbox{}^{\k]]}\D^{\th]}\Bigl]-(aa')^{D-4}i\Bigl[\mbox{}^\b\;_\b\D^{\si[\nu}\Bigl]\del^{[[\mu}\del'^{\th]} i\Bigl[\mbox{}^{\k]]}\D_\si\Bigl]\Bigl\}$ \Tstrut\Bstrut\\
\hline
2 & 1 & $2\k^2\del_\k\del'_\th\Bigl\{(aa')^{D-4}i\Bigl[\mbox{}^{\l[\mu}\D^\d\;_\d\Bigl]\del_\l\del'^{[[\nu} i\Bigl[\mbox{}^{\k]}\D^{\th]]}\Bigl]-(aa')^{D-4}i\Bigl[\mbox{}^{\rho[\mu}\D^\d\;_\d\Bigl]\del^{\k]}\del'^{[\nu} i\Bigl[\mbox{}_\rho\D^{\th]}\Bigl]\Bigl\}$ \Tstrut\Bstrut\\
\hline
2 & 2 & $-4\k^2\del_\k\del'_\th\Bigl\{(aa')^{D-4}i\Bigl[\mbox{}^{\l[\mu}\D^{\phi[[\nu}\Bigl]\del_\l\del'_\phi i\Bigl[\mbox{}^{\k]}\D^{\th]]}\Bigl]-(aa')^{D-4}i\Bigl[\mbox{}^{\rho[\mu}\D^{\phi[[\nu}\Bigl]\del^{\k]}\del'_\phi i\Bigl[\mbox{}_\rho\D^{\th]]}\Bigl]$ \Tstrut\Bstrut\\
 & & $+(aa')^{D-4}i\Bigl[\mbox{}^{\rho[\mu}\D^{\si[[\nu}\Bigl]\del^{\k]}\del'^{\th]]} i\Bigl[\mbox{}_\rho\D_\si\Bigl]-(aa')^{D-4}i\Bigl[\mbox{}^{\l[\mu}\D^{\si[[\nu}\Bigl]\del_\l\del'^{\th]]} i\Bigl[\mbox{}^{\k]}\D_\si\Bigl]\Bigr\}$ \Tstrut\Bstrut\\
\hline 
\end{tabular}
\caption{3-point results after naive index contractions.  \label{3naive}}
\end{table}
To complete the index contractions, substitutions for the graviton and photon propagator must be made.

\subsection{Graviton and Photon Propagator Substitutions}

To make substitutions for the graviton and photon propagators and completing the index contractions as clear as possible we will be using the new form of the graviton propagator (\ref{Gravprop2}) and also break the photon propagator up in a similar manner. To modify the orginal photon propagator (\ref{Photprop}) we can again add and subtract $B(x;x')$ from each term. Rearranging we find
\be
i\Bigl[\mbox{}_\mu\D_\nu\Bigr](x;x')=\e_{\mu\nu}aa'B(y)+\d_\mu^0\d_\nu^0\left[B(y)-C(y)\right] \; . 
\label{Photprop2}
\ee
Since we will be considering different parts of the propagators individually for the rest of this section it is necessary to take a moment to explain our notation. We will refer to the first, second, and third term of (\ref{Gravprop2}) with the subscripts $B$, $A$, and $C$ respectively. Likewise, for (\ref{Photprop2}) we will refer to the first and second terms with the subscripts $B$ and $C$ respectively.   

In the 3-point contribution it is always a product of the graviton and photon propagator that appear, so there are in general six combinations of terms that can arise $BB$, $AB$, $CB$, $BC$, $AC$, and $CC$, where the first letter stands for the part of the graviton propagator being considered and the second letter stands for the part of the photon propagator. However, it can be shown that the last two combinations do not need to be calculated since they are made of the product of two differences of scalar propagators. For these cases the conformal parts of the scalar propagators will cancel and what remains are only the infinite sums in each propagator, but these terms go like $\sim y^0$. Their degree of divergence is such that we can take $D=4$ immediately, and in this case all of the infinite series vanish. It is true that the A-type scalar propagator has an extra term going like $\sim y^{2-D/2}$; however, in the $D=4$ limit the infinite series multiplying this extra term will cause it to vanish too. So, in the end there are really only four combinations that need to be calculated. 

One other notational comment is needed before the calculation can continue. All four possible combinations of the parts of the propagators will result in terms of the same form, namely
\be
\#\del_\a\del'_\b\left\{(aa')^{D-4}i\D_x\del_\g\del'_\d(aa'i\D_y)\right\} \; ,
\label{genform}
\ee
where  $i\D_{x,y}$ is actually a difference of propagators for the cases $x=A,C$ or $y=C$. Since all of the terms will look almost identical, with only the propagators and indices changing, we can vastly simplify reporting our results by adopting a notation of only writing out the derivatives and their associated indices. These are the only parts needed because it is the indices that will make up the primary difference in each term, and the specific propagator combination can be denoted in the subscript of the vacuum polarization. In this notation (\ref{genform}) would take the simple form $\del_\a\del'_\b(\del_\g\del'_\d)$.

We are now ready to dive into the calculation. First we will perform the substitutions for the B part of the photon propagator with the B part of the graviton propagator. All of the terms for this portion will take the form 
\be
\#\del_\a\del'_\b\left\{(aa')^{D-4}B\del_\g\del'_\d(aa'B)\right\} \; .
\ee
The full result for this set of propagator pieces is
\bea
&&i\Bigl[\mbox{}^\mu\Pi_{\rm 3pt}^\nu\Bigl]_{\rm B,B}(x;x')=\f{(D^2-4D+2)}{(D-2)}\k^2\Bigl\{-\e^{\mu\nu}\del_\a\del'_\b(\del^\a\del'^\b) +\del^\nu\del'_\b(\del^\mu\del'^\b)-\del\cdot\del'(\del^\mu\del'^\nu) \nonumber \\
&&\hspace{1.4in}+\del_\a\del'^\mu(\del^\a\del'^\nu)\Bigr\} \nonumber \\
&&\hspace{0.2in}+\k^2\Bigl\{-2\e^{\mu\nu}\del\cdot\del'(\del\cdot\del')-\del\cdot\del'(\del^\nu\del'^\mu)+2\del^\mu\del'^\nu(\del\cdot\del')+\del_\a\del'^\mu(\del^\nu\del'^\a) \nonumber \\
&&\hspace{1.4in}-\e^{\mu\nu}\del_\a\del'_\b(\del^\b\del'^\a)+\del^\nu\del'_\b(\del^\b\del'^\mu)\Bigr\} \; .
\label{BBsub}
\eea

Next we compute the vacuum polarization from the $B$ part of the graviton propagator and the $C$ part of the photon propagator. These terms will all take the generic form
\be
\#\del_\a\del'_\b\left\{(aa')^{D-4}B\del_\g\del'_\d(aa'(B-C))\right\} \; .
\ee
The full result is
\bea
&&i\Bigl[\mbox{}^\mu\Pi_{\rm 3pt}^\nu\Bigl]_{\rm B,C}(x;x')=\f{2}{(D-2)}\k^2\Bigl\{\e^{\mu 0}\e^{\nu 0}\del_\a\del'_\b(\del^\a\del'^\b)-\e^{\nu 0}\del^0\del'_\b(\del^\mu\del'^\b)+\del^0\del'^0(\del^\mu\del'^\nu) \nonumber \\
&&\hspace{1.4in}-\e^{\mu 0}\del_\a\del'^0(\del^\a\del'^\nu)\Bigr\} \nonumber \\
&&\hspace{0.2in}+\k^2\Bigl\{-\e^{\mu\nu}\del^0\del'^0(\del\cdot\del')-\del^0\del'^0(\del^\nu\del'^\mu)+\e^{\mu\nu}\del_\a\del'^0(\del^\a\del'^0)+\e^{\mu\nu}\del_\a\del'_\b(\del^\a\del'^\b) \nonumber \\
&&\hspace{1.4in}+\e^{\mu\nu}\del^0\del'_\b(\del^0\del'^\b)+\e^{\mu 0}\del^\nu\del'^0(\del\cdot\del')+\e^{\mu 0}\del_\a\del'^0(\del^\nu\del'^\a)-\del^\nu\del'^0(\del^\mu\del'^0) \nonumber \\
&&\hspace{1.4in}-\del^\nu\del'_\b(\del^\mu\del'^\b)-\e^{\mu 0}\del^\nu\del'_\b(\del^0\del'^\b)-\e^{\mu 0}\e^{\nu 0}\del\cdot\del'(\del\cdot\del')-\e^{\mu 0}\e^{\nu 0}\del_\a\del'_\b(\del^\b\del'^\a) \nonumber \\
&&\hspace{1.4in}+\del\cdot\del'(\del^\mu\del'^\nu)+\e^{\nu 0}\del^0\del'^\mu(\del\cdot\del')+\e^{\nu 0}\del^0\del'_\b(\del^\b\del'^\mu)-\e^{\nu 0}\del_\a\del'^\mu(\del^\a\del'^0) \nonumber \\
&&\hspace{1.4in}+\e^{\nu 0}\del\cdot\del'(\del^\mu\del'^0)+\e^{\mu 0}\del\cdot\del'(\del^0\del'^\nu)-\del_\a\del'^\mu(\del^\a\del'^\nu)\hspace{0.2in}-\del^0\del'^\mu(\del^0\del'^\nu)\Bigr\} \; .
\label{BCsub}
\eea

When we consider the A part of the graviton propagator and the B part of the photon propagator the terms all have the form
\be
\#\del_\a\del'_\b\left\{(aa')^{D-4}(i\D_A-B)\del_\g\del'_\d(aa'B)\right\} \; ,
\ee
and the full result is
\bea
&&i\Bigl[\mbox{}^\mu\Pi_{\rm 3pt}^\nu\Bigl]_{\rm A,B}(x;x')=\f{(D-1)(D-2)}{(D-3)}\k^2\Bigl\{-\bar{\e}^{\mu\nu}\del_\a\del'_\b(\del^\a\del'^\b)+\bar{\del}^\nu\del'_\b(\del^\mu\del'^\b)-\bar{\del}\cdot\bar{\del}'(\del^\mu\del'^\nu) \nonumber \\
&&\hspace{1.4in}+\del_\a\bar{\del}'^\mu(\del^\a\del'^\nu)\Bigr\} \nonumber \\
&&\hspace{0.2in}+\k^2\Bigl\{-\bar{\e}^{\mu\nu}\del\cdot\del'(\bar{\del}\cdot\bar{\del}')-\del\cdot\del'(\bar{\del}^\nu\bar{\del}'^\mu)+\bar{\del}^\nu\del'^\mu(\bar{\del}\cdot\bar{\del}')+\bar{\del}_\a\del'^\mu(\bar{\del}^\nu\bar{\del}'^\a) \nonumber \\
&&\hspace{1.4in}-\e^{\mu\nu}\bar{\del}\cdot\bar{\del}'(\bar{\del}\cdot\bar{\del}')-\e^{\mu\nu}\bar{\del}_\a\bar{\del}'_\b(\bar{\del}^\b\bar{\del}'^\a)+\del^\nu\bar{\del}'^\mu(\bar{\del}\cdot\bar{\del}')+\del^\nu\bar{\del}'_\b(\bar{\del}^\b\bar{\del}'^\mu)\Bigr\} \nonumber \\
&&\hspace{0.2in}+\f{2}{(D-3)}\k^2\Bigl\{\del_\a\del'^\mu(\del^\a\bar{\del}'^\nu)-\e^{\mu\nu}\del_\a\bar{\del}'_\b(\del^\a\bar{\del}'^\b)-\del\cdot\del'(\del^\mu\bar{\del}'^\nu)+\del^\nu\bar{\del}'_\b(\del^\mu\bar{\del}'^\b) \nonumber \\
&&\hspace{1.4in}+\del^\nu\del'_\b(\bar{\del}^\mu\del'^\b)-\e^{\mu\nu}\bar{\del}_\a\del'_\b(\bar{\del}^\a\del'^\b)-\del\cdot\del'(\bar{\del}^\mu\del'^\nu)+\bar{\del}_\a\del'^\mu(\bar{\del}^\a\del'^\nu) \nonumber \\
&&\hspace{1.4in}+\del\cdot\del'(\bar{\del}^\mu\bar{\del}'^\nu)-\bar{\del}_\a\del'^\mu(\bar{\del}^\a\bar{\del}'^\nu)+\e^{\mu\nu}\bar{\del}_\a\bar{\del}'_\b(\bar{\del}^\a\bar{\del}'^\b)-\del^\nu\bar{\del}'_\b(\bar{\del}^\mu\bar{\del}'^\b)\Bigr\} \nonumber \\
&&\hspace{0.2in}+\f{(D-1)}{(D-3)}\k^2\Bigl\{\e^{\mu\nu}\del_\a\del'_\b(\del^\a\del'^\b)-\del^\nu\del'_\b(\del^\mu\del'^\b)+\del\cdot\del'(\del^\mu\del'^\nu)-\del_\a\del'^\mu(\del^\a\del'^\nu) \nonumber \\
&&\hspace{1.4in}+\bar{\e}^{\mu\nu}\del_\a\bar{\del}'_\b(\del^\a\bar{\del}'^\b)+\bar{\del}\cdot\bar{\del}'(\del^\mu\bar{\del}'^\nu)-\bar{\del}^\nu\bar{\del}'_\b(\del^\mu\bar{\del}'^\b)-\del_\a\bar{\del}'^\mu(\del^\a\bar{\del}'^\nu) \nonumber \\
&&\hspace{1.4in}-\bar{\del}^\nu\del'_\b(\bar{\del}^\mu\del'^\b)+\bar{\e}^{\mu\nu}\bar{\del}_\a\del'_\b(\bar{\del}^\a\del'^\b-\bar{\del}_\a\bar{\del}'^\mu(\bar{\del}^\a\del'^\nu)+\bar{\del}\cdot\bar{\del}'(\bar{\del}^\mu\del'^\nu)\Bigr\} \; .
\label{ABsub}
\eea

The last case is the $C$ part of the graviton propagator and the $B$ part of the photon propagator. These terms take the form
\be
\#\del_\a\del'_\b\left\{(aa')^{D-4}(C-B)\del_\g\del'_\d(aa'B)\right\} \; ,
\ee
and the full result is
\bea
&&i\Bigl[\mbox{}^\mu\Pi_{\rm 3pt}^\nu\Bigl]_{\rm C,B}(x;x')=2\f{(D-1)}{(D-2)}\k^2\Bigl\{-\e^{\mu 0}\del_\a\del'^0(\del^\a\del'^\nu)
+\e^{\nu 0}\e^{\mu 0}\del_\a\del'_\b(\del^\a\del'^\b)+\del^0\del'^0(\del^\mu\del'^\nu)
 \nonumber \\
&&\hspace{1.4in}-\e^{\nu 0}\del^0\del'_\b(\del^\mu\del'^\b)\Bigr\} \nonumber \\
&&\hspace{0.2in}+2\f{(D-3)}{(D-2)}\k^2\Bigl\{-\e^{\mu 0}\e^{\nu 0}\del\cdot\del'(\del^0\del'^0)+\e^{\nu 0}\del^0\del'^\mu(\del^0\del'^0)
-\e^{\mu\nu}\del^0\del'^0(\del^0\del'^0)+\e^{\mu 0}\del^\nu\del'^0(\del^0\del'^0)\Bigr\} \nonumber \\
&&\hspace{0.2in}+\f{2}{(D-2)(D-3)}\k^2\Bigl\{-\e^{\mu\nu}\del_\a\del'_\b(\del^\a\del'^\b)+\del^\nu\del'_\b(\del^\mu\del'^\b)
-\del\cdot\del'(\del^\mu\del'^\nu)+\del_\a\del'^\mu(\del^\a\del'^\nu) \nonumber \\
&& \hspace{1.4in}
-\del_\a\bar{\del}'^\mu(\del^\a\del'^\nu)+\bar{\e}^{\mu\nu}\del_\a\del'_\b(\del^\a\del'^\b)
+\bar{\del}\cdot\bar{\del}'(\del^\mu\del'^\nu)-\bar{\del}^\nu\del'_\b(\del^\mu\del'^\b) \nonumber \\
&& \hspace{1.4in}
-\del_\a\del'^\mu(\del^\a\bar{\del}'^\nu)+\e^{\mu\nu}\del_\a\bar{\del}'_\b(\del^\a\bar{\del}'^\b)
+\del\cdot\del'(\del^\mu\bar{\del}'^\nu)-\del^\nu\bar{\del}'_\b(\del^\mu\bar{\del}'^\b) \nonumber \\
&& \hspace{1.4in}
-\del^\nu\del'_\b(\bar{\del}^\mu\del'^\b)+\e^{\mu\nu}\bar{\del}_\a\del'_\b(\bar{\del}^\a\del'^\b)
+\del\cdot\del'(\bar{\del}^\mu\del'^\nu)-\bar{\del}_\a\del'^\mu(\bar{\del}^\a\del'^\nu) \nonumber \\
&& \hspace{1.4in}
-\del\cdot\del'(\bar{\del}^\mu\bar{\del}'^\nu)+\del_\a\bar{\del}'^\mu(\del^\a\bar{\del}'^\nu)
+\bar{\del}^\nu\del'_\b(\bar{\del}^\mu\del'^\b)+\bar{\del}_\a\del'^\mu(\bar{\del}^\a\bar{\del}'^\nu) \nonumber \\
&& \hspace{1.4in}
-\bar{\del}\cdot\bar{\del}'(\del^\mu\bar{\del}'^\nu)-\bar{\e}^{\mu\nu}\bar{\del}_\a\del'_\b(\bar{\del}^\a\del'^\b)
-\e^{\mu\nu}\bar{\del}_\a\bar{\del}'_\b(\bar{\del}^\a\bar{\del}'^\b)+\bar{\del}^\nu\bar{\del}'_\b(\del^\mu\bar{\del}'^\b) \nonumber \\
&& \hspace{1.4in}
+\bar{\del}_\a\bar{\del}'^\mu(\bar{\del}^\a\del'^\nu)+\del^\nu\bar{\del}'_\b(\bar{\del}^\mu\bar{\del}'^\b)
-\bar{\e}^{\mu\nu}\del_\a\bar{\del}'_\b(\del^\a\bar{\del}'^\b)-\bar{\del}\cdot\bar{\del}'(\bar{\del}^\mu\del'^\nu)\Bigl\} \nonumber \\
&&\hspace{0.2in}+\f{2}{(D-2)}\k^2\Bigl\{-\e^{\nu 0}\del_\a\del'^\mu(\del^\a\del'^0)-\e^{\mu 0}\del^\nu\del'_\b(\del^0\del'^\b)
+\e^{\mu\nu}\del_\a\del'^0(\del^\a\del'^0)+\e^{\mu\nu}\del^0\del'_\b(\del^0\del'^\b) \nonumber \\
&& \hspace{1.4in}
+\e^{\nu 0}\del\cdot\del'(\del^\mu\del'^0)+\e^{\mu 0}\del\cdot\del'(\del^0\del'^\nu)
-\del^\nu\del'^0(\del^\mu\del'^0)-\del^0\del'^\mu(\del^0\del'^\nu) \nonumber \\
&& \hspace{1.4in}
-\e^{\mu 0}\del\cdot\del'(\del^0\bar{\del}'^\nu)-\e^{\nu 0}\del\cdot\del'(\bar{\del}^\mu\del'^0)
+\e^{\mu 0}\del_\a\del'^0(\del^\a\bar{\del}'^\nu)+\e^{\nu 0}\del_\a\bar{\del}'^\mu(\del^\a\del'^0) \nonumber \\
&& \hspace{1.4in}
+\e^{\mu 0}\bar{\del}^\nu\del'_\b(\del^0\del'^\b)+\e^{\nu 0}\del^0\del'_\b(\bar{\del}^\mu\del'^\b)
+\del^0\del'^\mu(\del^0\bar{\del}'^\nu)+\e^{\nu 0}\bar{\del}_\a\del'^\mu(\bar{\del}^\a\del'^0) \nonumber \\
&& \hspace{1.4in}
-\del^0\del'^0(\del^\mu\bar{\del}'^\nu)-\e^{\nu 0}\bar{\del}\cdot\bar{\del}'(\del^\mu\del'^0)
-\bar{\e}^{\mu\nu}\del^0\del'_\b(\del^0\del'^\b)-\e^{\nu 0}\e^{\mu 0}\bar{\del}_\a\del'_\b(\bar{\del}^\a\del'^\b) \nonumber \\
&& \hspace{1.4in}
-\e^{\mu\nu}\del^0\bar{\del}'_\b(\del^0\bar{\del}'^\b)-\e^{\mu\nu}\bar{\del}_\a\del'^0(\bar{\del}^\a\del'^0) 
+\e^{\nu 0}\del^0\bar{\del}'_\b(\del^\mu\bar{\del}'^\b)+\bar{\del}^\nu\del'^0(\del^\mu\del'^0) \nonumber \\
&& \hspace{1.4in}
+\del^0\bar{\del}'^\mu(\del^0\del'^\nu)+\e^{\mu 0}\bar{\del}_\a\del'^0(\bar{\del}^\a\del'^\nu) 
+\e^{\mu 0}\del^\nu\bar{\del}'_\b(\del^0\bar{\del}'^\b)+\del^\nu\del'^0(\bar{\del}^\mu\del'^0) \nonumber \\
&& \hspace{1.4in}
-\e^{\mu 0}\e^{\nu 0}\del_\a\bar{\del}'_\b(\del^\a\bar{\del}'^\b)-\bar{\e}^{\mu\nu}\del_\a\del'^0(\del^\a\del'^0) 
-\e^{\mu 0}\bar{\del}\cdot\bar{\del}'(\del^0\del'^\nu) \nonumber \\
&&\hspace{1.4in}-\del^0\del'^0(\bar{\del}^\mu\del'^\nu)\Bigr\} \; .
\label{CBsub}
\eea

\subsection{Finding the 3-Point Structure Functions}

Recalling the procedure outlined in section \ref{4sf} we will now find the 3-point contributions to $F$ and $G$ in much the same way. The process will be a little more labor intensive since we no longer have a delta function on each term to assist in pulling out internal derivatives; there is also the added complication of having two internal derivatives instead of one. These changes will be accounted for as follows:

First we can still use the same identities (\ref{00id}) and (\ref{ijid}) as our set of equations for finding $F_3$ and $G_4$. Next we notice that in the absence of a delta function we can no longer simply change $\del\rightarrow-\del'$, instead we have to carefully account for feed down terms that will arise from extracting derivatives. All of the 3-point terms take the form of (\ref{genform}), thus there are only four possible combinations of internal derivatives we will encounter, they are $\del_i\del'_j$, $\del_0\del'_i$, $\del_i\del'_0$, and $\del_0\del'_0$. Extracting these sets of derivatives will always result in the same feed down terms regardless of the propagators involved. A slight modification is needed for the A-type propagator, but in general extracting these derivatives results in the following identities
\bea
(aa')^{D-4}i\D_x\del_0\del'_i(aa'i\D_y) & = & \del_0\del'_i\left[(aa')^{D-3}I^2[i\D_xi\ddot{\D}_y]\right]+\del'_i\Bigl\{H(aa')^{D-3}a\left[(D-2)I^2[i\dot{\D}_xi\dot{\D}_y]\right. \nonumber \\
&&\hspace{0.5in}\left.-(D-4)I[i\D_xi\dot{\D}_y]\right]\Bigr\} \; , \label{0ipid}\\
&& \nonumber \\
(aa')^{D-4}i\D_x\del_i\del'_0(aa'i\D_y)& = &\del_i\del'_0\left[(aa')^{D-3}I^2[i\D_xi\ddot{\D}_y]\right]+\del_i\Bigl\{H(aa')^{D-3}a'\left[(D-2)I^2[i\dot{\D}_xi\dot{\D}_y]\right. \nonumber \\
&&\hspace{0.5in}\left.-(D-4)I[i\D_xi\dot{\D}_y]\right]\Bigr\} \; , \label{i0pid}\\
&& \nonumber \\
(aa')^{D-4}i\D_x\del_i\del'_j(aa'i\D_y) & = & \del_i\del'_j\left[(aa')^{D-3}I^2[i\D_xi\ddot{\D}_y]\right]-2\e_{ij}\left[H^2(aa')^{D-2}I[i\dot{\D}_xi\dot{\D}_y]\right] \; , \label{ijpid} \\
&& \nonumber \\
(aa')^{D-4}i\D_x\del_0\del'_0(aa'i\D_y)& = &\del_0\del'_0\left[(aa')^{D-3}I^2[i\D_xi\ddot{\D}_y]\right]+H^2(aa')^{D-2}\Bigl\{(D-3)(D-5)I[i\D_xi\dot{\D}_y] \nonumber \\
&&\hspace{0.5in} -(D-3)(D-1)I^2[i\dot{\D}_xi\dot{\D}_y]+(2-y)I[i\dot{\D}_xi\dot{\D}_y]+i\D_xi\D_y\Bigr\} \nonumber \\
&&\hspace{0.5in} +(\del_0a'+\del'_0a)\Bigl\{(aa')^{D-3}H\left[(D-2)I^2[i\dot{\D}_xi\dot{\D}_y]\right. \nonumber \\
&&\hspace{0.5in} \left.-(D-4)I[i\D_xi\dot{\D}_y]\right]\Bigr\} \; , \label{00pid}
\eea
where $I$ represents and an indefinite integral with respect to $y$, and a dot over the propagators represents a derivative with respect to y\footnote{One might worry about local delta function contributions from terms of the form $\del_0\del'_0B$, but actually these contributions are fully accounted for in delta functions arising from feed down terms of the form $\del_0\del'_0\left[(aa')^{D-3}I^2[i\D_xi\ddot{\D}_y]\right]$.} . The A-type propagator is unique in that it is not just a function of $y$, but also contains a de Sitter breaking piece. It can be rewritten in the form
\be
i\D_A(x;x')=A(y)+ku \; ,
\label{speciala}
\ee
where $k=\f{H^{D-2}}{(4\pi)^{D/2}}\f{\G(D-1)}{\G(D/2)}$. In this form we can see that identities (\ref{0ipid}-\ref{00pid}) will miss the feed down terms that arise when one of the internal derivatives act on the second term in (\ref{speciala}). To account for this the following terms need to be inserted in each of the above identities when $i\D_x=(i\D_A-B)$
\bea
&&(aa')^{D-4}ku\del_0\del'_i(aa'B)\rightarrow -\del'_i\left[Ha(aa')^{D-3}kB\right] \label{ku0i}\\
&& \nonumber \\
&&(aa')^{D-4}ku\del_i\del'_0(aa'B)\rightarrow -\del_i\left[Ha'(aa')^{D-3}kB\right] \\
&& \nonumber \\
&&(aa')^{D-4}ku\del_i\del'_j(aa'B)\rightarrow 0 \\
&& \nonumber \\
&&(aa')^{D-4}ku\del_0\del'_0(aa'B)\rightarrow -(\del_0a'+\del'_0a)\times\left[H(aa')^{D-3}kB\right] \nonumber \\
&&\hspace{2in}+2H^2(D-4)(aa')^{D-2}kB \label{ku00}
\eea

We are now ready to apply the contractions in identities (\ref{00id}) and (\ref{ijid}) and then use the substitutions (\ref{0ipid}-\ref{00pid}), making use of (\ref{ku0i}-\ref{ku00}) where appropriate, to find $F_3$ and $G_3$. Again, we will work through an example for the reader and then state the final result for all of the terms. 

We consider all terms with the coefficient $\f{(D^2-4D+2)}{(D-2)}\k^2$ in (\ref{BBsub}), since these form the smallest set of transverse terms. Performing the first contraction with $\d^0_\mu\d^0_\nu$ results in
\bea
\lefteqn{i\Bigl[\mbox{}^0\Pi^0_{\rm 3pt,ex}\Bigr]_{B,B}=\f{(D^2-4D+2)}{(D-2)}\k^2\Bigl\{\del_\a\del'_\b\left[(aa')^{D-4}B\del^\a\del'^\b(aa'B)\right]} \nonumber \\
&&+\del^0\del'_\b\left[(aa')^{D-4}B\del^0\del'^\b(aa'B)\right]-\del\cdot\del'\left[(aa')^{D-4}B\del^0\del'^0(aa'B)\right] \nonumber \\
&&+\del_\a\del'^0\left[(aa')^{D-4}B\del^\a\del'^0(aa'B)\right]\Bigr\} \nonumber \\
&=&\f{(D^2-4D+2)}{(D-2)}\k^2\Bigl\{\del_i\del'_j\left[(aa')^{D-4}B\del_i\del'_j(aa'B)\right] \nonumber \\
&&+\nabla^2\left[(aa')^{D-4}B\del_0\del'_0(aa'B)\right]\Bigr\}
\eea
where we have taken the 3+1 decomposition in going from the first line to the second. Now we can make the appropriate substitutions using identities (\ref{0ipid}-\ref{00pid}) and compare with (\ref{00id}) to find
\bea
\lefteqn{F_{\rm 3,ex}(x;x')=\f{(D^2-4D+2)}{(D-2)}\k^2\Bigl\{(\del_0\del'_0+\nabla^2)\left[(aa')^{D-3}I^2[B\ddot{B}]\right]} \nonumber \\
&&+H(\del_0a'+\del'_0a)\times\left[(aa')^{D-3}\left((D-2)I^2[\dot{B}^2]-(D-4)I[B\dot{B}]\right)\right] \nonumber \\
&&+H^2(aa')^{D-2}\left[B^2+(4-y)I[{\dot{B}}^2]-(D-1)(D-3)I^2[{\dot{B}}^2]+(D-3)(D-5)I[B\dot{B}]\right]\Bigr\} \; .
\eea
To find the companion $G_{\rm 3,ex}$ we contract $\d^i_\mu\d^j_\nu(i\neq j)$ with the same set of terms in (\ref{BBsub})
\bea
\lefteqn{i\Bigl[\mbox{}^i\Pi^j_{\rm 3pt,ex}\Bigr]_{B,B}=\f{(D^2-4D+2)}{(D-2)}\k^2\Bigl\{0+\del^j\del'_\b\left[(aa')^{D-4}B\del^i\del'^\b(aa'B)\right]} \nonumber \\
&&-\del\cdot\del'\left[(aa')^{D-4}B\del^i\del'^j(aa'B)\right]+\del_\a\del'^i\left[(aa')^{D-4}B\del^\a\del'^j(aa'B)\right]\Bigr\} \nonumber \\
&=&\f{(D^2-4D+2)}{(D-2)}\k^2\Bigl\{-\del_j\del'_0\left[(aa')^{D-4}B\del_i\del'_0(aa'B)\right]+\del_j\del'_k\left[(aa')^{D-4}B\del_i\del'_k(aa'B)\right] \nonumber \\
&&+\del_0\del'_0\left[(aa')^{D-4}B\del_i\del'_j(aa'B)\right]+\nabla^2\left[(aa')^{D-4}B\del_i\del'_j(aa'B)\right] \nonumber \\
&&-\del_0\del'_i\left[(aa')^{D-4}B\del_0\del'_j(aa'B)\right]+\del_k\del'_i\left[(aa')^{D-4}B\del_k\del'_j(aa'B)\right] \; ,
\label{gcont}
\eea
where again we have taken the 3+1 decomposition in going from the first line to the second. Applying identities (\ref{0ipid}-\ref{00pid}), (\ref{gcont}) can be rewritten in the form 
\bea
\lefteqn{i\Bigl[\mbox{}^i\Pi^j_{\rm 3pt,ex}\Bigr]_{B,B}=-\f{(D^2-4D+2)}{(D-2)}\k^2\del'_i\del_j\Bigl\{(\nabla^2-\del_0\del'_0-{\del'_0}^2-{\del_0}^2)\times} \nonumber \\
&&\left[(aa')^{D-3}I^2[B\ddot{B}]\right] \nonumber \\
&&-\del'_0\left[H(aa')^{D-3}a'\left(I[B\dot{B}]-(D-3)I^2[B\ddot{B}]+I^2[\dot{B}]\right)\right] \nonumber \\
&&-\del_0\left[H(aa')^{D-3}a\left(I[B\dot{B}]-(D-3)I^2[B\ddot{B}]+I^2[\dot{B}]\right)\right] \nonumber \\
&&+4H^2(aa')^{D-2}I[{\dot{B}}^2]\Bigr\} \; .
\eea
Substituting $F_{\rm 3,ex}$ into (\ref{ijid}) we find 
\bea
\lefteqn{G_{\rm 3,ex}(x;x')=-\f{(D^2-4D+2)}{(D-2)}\k^2\left\{H^2(aa')^{D-2}\left({B}^2+(D-3)(D-5)I[B\dot{B}]\right.\right.} \nonumber \\
&&\left.-(D-1)(D-3)I^2[{\dot{B}}^2]-yI[{\dot{B}}^2]\right) \nonumber \\
&&+H(\del_0a'+\del'_0a)\times\left[(aa')^{D-3}\left((D-2)I^2[\dot{B}^2]-(D-4)I[B\dot{B}]\right)\right] \nonumber \\
&&+H(\del_0a+\del'_0a')\times\left[(aa')^{D-3}\left((D-2)I^2[\dot{B}^2]-(D-4)I[B\dot{B}]\right)\right] \nonumber \\
&&\left.+(\del_0+\del'_0)^2\left[(aa')^{D-3}I^2[B\ddot{B}]\right]\right\} \; .
\eea
This concludes the 3-point example; it should be clear how to proceed with the rest of the 3-point contribution.

We now present the results for the scalar structure functions organized according to propagator combination. From (\ref{BBsub}) we find 
\bea
\lefteqn{F_{BB}(x;x')=\f{(D^2-D-4)}{(D-2)}\k^2\Bigl\{H^2(aa')^{D-2}\left[{B}^2+(D-3)(D-5)I[B\dot{B}]\right.} \nonumber \\
&&\left.-(D-1)(D-3)I^2[{\dot{B}}^2]\right]+\left(\nabla^2+\del_0\del'_0\right)\times\left[(aa')^{D-3}I^2[B\ddot{B}]\right]\nonumber \\
&&+H\left(\del_0a'+\del'_0a\right)\times\left[(aa')^{D-3}\left((D-2)I^2[\dot{B}^2]-(D-4)I[B\dot{B}]\right)\right]\Bigr\} \nonumber \\
&&+\f{\k^2H^2}{(D-2)}(aa')^{D-2}\left[4D(2D-5)-(D^2-D-4)y\right]I[\dot{B}^2] \; ,\label{FBB} \\
&& \nonumber \\
\lefteqn{G_{BB}(x;x')=-\f{D(D-3)}{(D-2)}\k^2\Bigl\{H^2(aa')^{D-2}\left[{B}^2+(D-3)(D-5)I[B\dot{B}]\right.} \nonumber \\
&&\left.-(D-1)(D-3)I^2[{\dot{B}}^2]-yI[{\dot{B}}^2]\right]\Bigr\}\nonumber \\
&&-\f{(D^2-4D+2)}{(D-2)}\k^2\Bigl\{H\left(\del_0a'+\del'_0a\right)\times\left[(aa')^{D-3}\left((D-2)I^2[{\dot{B}}^2]-(D-4)I[B\dot{B}]\right)\right] \nonumber \\
&&+H\left(\del_0a+\del'_0a'\right)\times\left[(aa')^{D-3}\left((D-2)I^2[{\dot{B}}^2]-(D-4)I[B\dot{B}]\right)\right] \nonumber \\
&&+(\del_0+\del'_0)^2\left[(aa')^{D-3}I^2[B\ddot{B}]\right]\Bigr\} \; . \label{GBB}
\eea

From (\ref{BCsub}) we find
\bea
\lefteqn{F_{BC}(x;x')=-\f{\k^2}{(D-2)}\Bigl\{2(D^2-D-4)H^2(aa')^{D-2}I[\dot{B}(\dot{B}-\dot{C})]} \nonumber \\
&&\hspace{1in}+(3D-8)(aa')^{D-3}\nabla^2I^2[B(\ddot{B}-\ddot{C})]\Bigr\} \; , \label{FBC} \\
&& \nonumber \\
\lefteqn{G_{BC}(x;x')=2\k^2\f{(D-3)}{(D-2)}\Bigl\{DH^2(aa')^{D-2}I[\dot{B}(\dot{B}-\dot{C})]} \nonumber \\
&&\hspace{1in}+\left(\nabla^2+\del_0\del'_0\right)\times\left[(aa')^{D-3}I^2[B(\ddot{B}-\ddot{C})]\right]\Bigr\} \; . \label{GBC}
\eea

From (\ref{ABsub}) we have 
\bea
\lefteqn{F_{AB}(x;x')=\k^2\Bigl\{(D-1)H^2(aa')^{D-2}\left[(i\D_A-B)B+(D-3)(D-5)I[(i\D_A-B)\dot{B}]\right.} \nonumber \\
&&\left.-(D-1)(D-3)I^2[(i\dot{\D}_A-\dot{B})\dot{B}]+(4-y)I[(i\dot{\D}_A-\dot{B})\dot{B}]\right] \nonumber \\
&&+\left(\nabla^2+(D-1)\del_0\del'_0\right)\times\left[(aa')^{D-3}I^2[(i\D_A-B)\ddot{B}]\right] \nonumber \\
&&+(D-1)H(\del_0a'+\del'_0a)\times\left[(aa')^{D-3}\left((D-2)I^2[(i\dot{\D}_A-\dot{B})\dot{B}]-(D-4)I[(i\D_A-B)\dot{B}]\right)\right]\Bigr\} \nonumber \\
&&+(D-1)k\k^2\Bigl\{2(D-4)H^2(aa')^{D-2}B-H(\del_0a'+\del'_0a)\left[(aa')^{D-3}B\right]\Bigr\} \; , \label{FAB} \\
&& \nonumber \\
\lefteqn{G_{AB}(x;x')=-\k^2\Bigl\{(D-1)H^2(aa')^{D-2}\left[(i\D_A-B)B+(D-3)(D-5)I[(i\D_A-B)\dot{B}]\right.} \nonumber \\
&&\left.-(D-1)(D-3)I^2[(i\dot{\D}_A-\dot{B})\dot{B}]-\left(4\f{(D-4)}{(D-3)}+y\right)I[(i\dot{\D}_A-\dot{B})\dot{B}]\right] \nonumber \\
&&+\left[-\f{(D^2-4D+1)}{(D-3)}\nabla^2+(D-2)(\del_0^2+{\del'_0}^2)+(D-1)\del_0\del'_0\right]\times\left[(aa')^{D-3}I^2[(i\D_A-B)\ddot{B}]\right] \nonumber \\
&&+(D-1)H(\del_0a'+\del'_0a)\times\left[(aa')^{D-3}\left((D-2)I^2[(i\dot{\D}_A-\dot{B})\dot{B}]-(D-4)I[(i\D_A-B)\dot{B}]\right)\right] \nonumber \\
&&+(D-2)H(\del_0a+\del'_0a')\times\left[(aa')^{D-3}\left((D-2)I^2[(i\dot{\D}_A-\dot{B})\dot{B}]-(D-4)I[(i\D_A-B)\dot{B}]\right)\right]\Bigl\} \nonumber \\
&&-k\k^2\Bigl\{2(D-4)(D-1)H^2(aa')^{D-2}B-(D-1)H(\del_0a'+\del'_0a)\times\left[(aa')^{D-3}B\right] \nonumber \\
&&-(D-2)H(\del_0a+\del'_0a')\times\left[(aa')^{D-3}B\right]\Bigl\} \; . \label{GAB}
\eea
And lastly, from (\ref{CBsub}) we have
\bea
\lefteqn{F_{CB}(x;x')=2\f{(D-3)}{(D-2)}\k^2\Bigl\{H^2(aa')^{D-2}\left[(C-B)B+(D-3)(D-5)I[(C-B)\dot{B}]\right.} \nonumber \\
&&\left.-(D-1)(D-3)I^2[(\dot{C}-\dot{B})\dot{B}]+(4-y)I[(\dot{C}-\dot{B})\dot{B}]\right] \nonumber \\
&&+H(\del_0a'+\del'_0a)\times\left[(aa')^{D-3}\left((D-2)I^2[(\dot{C}-\dot{B})\dot{B}]-(D-4)I[(C-B)\dot{B}]\right)\right] \nonumber \\
&&+(\nabla^2+\del_0\del'_0)\left[(aa')^{D-3}I^2[(C-B)\ddot{B}]\right]\Bigr\} \; , \label{FCB} \\
&& \nonumber \\
\lefteqn{G_{CB}(x;x')=-\f{2\k^2}{(D-2)}\Bigl\{H^2(aa')^{D-2}\left[(D-3)(C-B)B+(D-3)^2(D-5)I[(C-B)\dot{B}]\right.} \nonumber \\
&&\left.-(D-3)^2(D-1)I^2[(\dot{C}-\dot{B})\dot{B}]+\left(4\f{(D-2)(D-4)}{(D-3)}-(D-3)y\right)I[(\dot{C}-\dot{B})\dot{B}]\right] \nonumber \\
&&+\left[\f{(D-4)(D-2)}{(D-3)}\nabla^2+2(D-3)\del_0\del'_0-\del_0^2-{\del'_0}^2\right]\times\left[(aa')^{D-3}I^2[(C-B)\ddot{B}]\right] \nonumber \\
&&+(D-3)H(\del_0a'+\del'_0a)\times\left[(aa')^{D-3}\left((D-2)I^2[(\dot{C}-\dot{B})\ddot{B}]-(D-4)I[(C-B)\dot{B}]\right)\right] \nonumber \\
&&-H(\del_0a+\del'_0a')\times\left[(aa')^{D-3}\left((D-2)I^2[(\dot{C}-\dot{B})\ddot{B}]-(D-4)I[(C-B)\dot{B}]\right)\right] \Bigr\} \; . \label{GCB}
\eea
Because of the different combinations of propagators appearing in each set of $F$ and $G$ it was not useful or enlightening to combine them here. Instead, once they are renormalized in the next section they are combined easily.

%\subsection{Delta Function Contributions}
%
%There is actually one more set of structure functions coming from delta functions produced when two time derivatives act on $B$. Recall the B-type propagator equation (\ref{bpropeq})
%\be
%\left[\square-(D-2)H^2\right]B(x;x')=\f{i\d^D(x-x')}{\sqrt{-g}} \; .
%\ee
%We need to accurately account for this delta function when extracting the internal derivatives. In particular we need to account for delta functions arising from terms of the form $\del_0\del'_0B$. To do this we can apply the same identities (\ref{00id}) and (\ref{ijid}) to (\ref{BBsub}), (\ref{BCsub}), (\ref{ABsub}), and (\ref{CBsub}) only keeping $\del_0\del'_0B$ terms that will contribute a delta function. When all of these terms have been added together we find another set of structure functions both proportional to a delta function
%\bea
%F_{3\d}(x;x')&=&\left\{(D-1)i\D_A(0)+2\f{(D-3)}{(D-2)}C(0)\right\}\k^2a^{D-4}i\d^D(x-x') \label{3fdelta} \\
%&& \nonumber \\
%G_{3\d}(x;x')&=&\left\{2B(0)-(D-1)i\D_A(0)-2\f{(D-3)}{(D-2)}C(0)\right\}\k^2a^{D-4}i\d^D(x-x') \; . \label{3gdelta}
%\eea
%In this form $F_{3\d}$ and $G_{3\d}$ will combine with the 4-point structure function contributions (\ref{f4}) and (\ref{g4}). We have now found the 3-point structure functions contributing to the vacuum polarization, now it only remains to renormalize these results.

%---------------------------------------------------------------------------------------------------------------------------------------------------------------%
\section{Renormalization}\label{Renorm}

This section is devoted to renormalizing the 3 and 4-point contributions to $F$ and $G$. First, all of the 3-point contributions must be put in the same form so as to be easily combined. We can then localize the ultraviolet divergent pieces, and combine them with the 4-point divergences. By reading off the correct counterterm coefficients we remove all divergences, and are left with the fully renormalized structure functions of the vacuum polarization.

\subsection{Converting to Functions of $\D x$}

The expressions we derived in section~\ref{3point} contain many 
indefinite integrals of products of derivatives of the three 
propagator functions $i\D_A(y)$, $B(y)$ and $C(y)$. Each of these
products consists of a few powers of $y(x;x')$ which are singular 
at coincidence (${x'}^{\mu} = x^{\mu}$) and whose coefficients are 
nonzero for $D=4$, plus an infinite series of less and less singular 
powers of $y$ whose coefficients vanish for $D=4$. Because the vacuum 
polarization is used inside the 4-dimensional integral of the 
quantum-corrected Maxwell equation (\ref{QMax}), the only terms which 
require dimensional regularization are those which are at least as
singular as $1/y^2$. Any less singular term can be evaluated for
$D=4$, at which point most of the tedious infinite series contributions
vanish. Recalling as well that $y(x;x') = H^2 a a' \Delta x^2$, we
can make the following simplifications:

%There are many different products of scalar propagators and their derivatives and antiderivatives in the $F$'s and $G$'s of section \ref{3point}. All of these terms will contain infinite series, thus it is pertinent to understand how many terms need to be kept for each combination. Since all of the infinite series will go to zero in the $D=4$ limit it is only necessary to keep those terms containing powers of $y^{-2}$ and higher in the propagator products. We found that in no case is anything more than the first term in the infinite series ever necessary to keep. In general terms that go like $1/\D x^2$ and higher are not divergent, and we can take the limit $D=4$ immediately. We shall now list identities that can be used to transform the results of section \ref{3point} into functions of $\D x$, being careful to only keep terms that will contribute in the limit $D=4$. 

For $F_{BB}$ and $G_{BB}$ we use the identities
\bea
{B}^2& \rightarrow &\f{\G^2\left(\f{D}{2}-1\right)}{16\pi^D(aa')^{D-2}\D x^{2D-4}} \\
I[B\dot{B}]& \rightarrow &\f{\G^2\left(\f{D}{2}-1\right)}{32\pi^D(aa')^{D-2}\D x^{2D-4}} \\
I[{\dot{B}}^2]& \rightarrow &-\f{(D-2)^2}{(D-1)}\f{\G^2\left(\f{D}{2}-1\right)}{64\pi^D(aa')^{D-1}H^2\D x^{2D-2}}-\f{(D-2)}{256\pi^D}\f{\G^2\left(\f{D}{2}-1\right)(D-4)}{(aa')^{D-2}\D x^{2D-4}} \\
I^2[{\dot{B}}^2]& \rightarrow &\f{(D-2)}{(D-1)}\f{\G^2\left(\f{D}{2}-1\right)}{64\pi^D(aa')^{D-2}\D x^{2D-4}} \\
I^3[{\dot{B}}^2]& \rightarrow &\f{H^2}{96\pi^4}\f{1}{aa'\D x^2} \label{i3bpbp}\\
I^2[B\ddot{B}]& \rightarrow &\f{D}{(D-1)}\f{\G^2\left(\f{D}{2}-1\right)}{64\pi^D(aa')^{D-2}\D x^{2D-4}} \label{i2bbpp}
\eea
where we note that for this set of identities only (\ref{i3bpbp}) could be put in the $D=4$ limit. For $F_{BC}$ and $G_{BC}$ most of the terms will go to zero and there is only one identity needed
\be
I[\dot{B}(\dot{B}-\dot{C}))]\rightarrow-\f{(D-4)}{128}\f{\G^2\left(\f{D}{2}-1\right)}{\pi^D(aa')^{D-2}\D x^{2D-4}} \; .
\ee

For $F_{AB}$ and $G_{AB}$ we find that we can take $D=4$ in most of the terms
\bea
B(i\D_A-B)& \rightarrow &-\f{H^2}{32\pi^4}\left[\f{1}{2}+\ln\left(\f{1}{4}H^2\D x^2\right)\right]\f{1}{aa'\D x^2} \\
I[(i\D_A-B)\dot{B}]& \rightarrow &-\f{H^2}{32\pi^4}\left[\f{3}{2}+\ln\left(\f{1}{4}H^2\D x^2\right)\right]\f{1}{aa'\D x^2} \\
I[(i\dot{\D}_A-\dot{B})\dot{B}]& \rightarrow &-\f{(D-2)}{2}\f{\G^2\left(\f{D}{2}-1\right)}{64\pi^D}\f{1}{(aa')^{D-2}\D x^{2D-4}} \label{iabp} \\
I^2[(i\dot{\D}_A-\dot{B})\dot{B}]& \rightarrow &\f{H^2}{64\pi^4}\f{1}{aa'\D x^2} \\
I^3[(i\dot{\D}_A-\dot{B})\dot{B}]& \rightarrow &\f{H^4}{64\pi^4}\left[\ln\left(\f{1}{4}H^2\D x^2\right)+u\right] \\
I^2[(i\D_A-B)\ddot{B}]& \rightarrow &-\f{H^2}{32\pi^4}\left[2+\ln\left(\f{1}{4}H^2\D x^2\right)\right]\f{1}{aa'\D x^2} \\
kB& \rightarrow &\f{H^2}{32\pi^4}\f{1}{aa'\D x^2}
\eea
Notice that (\ref{iabp}) is the only identity that needs to be kept in $D$ dimensions. Lastly for $F_{CB}$ and $G_{CB}$ there is again only one relevant identity since most of the terms will be zero
\be
I[(\dot{C}-\dot{B})\dot{B}]\rightarrow\f{(D-4)}{128}\f{\G^2\left(\f{D}{2}-1\right)}{\pi^D(aa')^{D-2}\D x^{2D-4}} \; .
\ee

\subsection{Finding the Finite and Divergent Parts of F and G}

It is easiest to renormalize the structure functions term by term; thus we will demonstrate the procedure for one term and the reader can extrapolate from the example to derive the rest of the terms. First, it is useful to note the necessary identities for renormalization. The terms proportional to $1/\D x^2$ are already integrable and do not need to be renormalized. There are also terms proportional $1/\D x^{2D-2}$ and $1/\D x^{2D-4}$; these will need to be renormalized. We use dimensional regulation to partially integrate these terms until they are integrable 
\bea
\f{1}{\D x^{2D-2}}&=&\f{\del^4}{4(D-2)^2(D-3)(D-4)}\f{1}{\D x^{2D-6}} \; , \label{id1} \\
\f{1}{\D x^{2D-4}}&=&\f{\del^2}{2(D-3)(D-4)}\f{1}{\D x^{2D-6}} \; , \label{id2}
\eea
but it is clear that these identities contain a divergence in the factors of $\f{1}{(D-4)}$. We can localize the divergence by adding zero in the form
\be
\del^2\f{1}{\D x^{D-2}}=\f{i4\pi^{D/2}}{\G\left(\f{D}{2}\!-\!1\right)}\d^D(x-x') \; .
\label{zero}
\ee
Then by adding (\ref{zero}) to the divergent parts of (\ref{id1}) and (\ref{id2}) we find 
\bea
\f{\del^2}{(D-4)}\f{1}{\D x^{2D-6}}&=&\f{i4\pi^{D/2}}{\G\left(\f{D}{2}\!-\!1\right)}\f{\mu^{D-4}\d^D(x-x')}{(D-4)}-\f{\del^2}{(D-4)}\left\{\f{1}{\D x^{2D-6}}-\f{\mu^{D-4}}{\D x^{D-2}}\right\} \nonumber \\
&=&\f{i4\pi^{D/2}}{\G\left(\f{D}{2}\!-\!1\right)}\f{\mu^{D-4}\d^D(x-x')}{(D-4)}-\f{\del^2}{2}\left[\f{\ln(\mu^2\D x^2)}{\D x^2}\right]+\mathcal{O}(D-4) \; ,  \label{id3}
\eea
where the factor of $\mu$ is added for dimensional consistency. We may now begin our example:

We consider the eighth term in $F_{BB}$ 
\be
F_{\rm ex}(x;x')=\f{(D^2-D-4)}{(D-2)}\k^2(\nabla^2+\del_0\del'_0)\left[(aa')^{D-3}I^2[B\ddot{B}]\right] \; .
\ee
Applying identity (\ref{i2bbpp}) we find
\be
F_{\rm ex}(x;x')=\f{(D^2-D-4)}{(D-2)}\k^2(\nabla^2+\del_0\del'_0)\left[\f{D}{64\pi^D(D-1)}\f{\G^2(\f{D}{2}-1)}{aa'\D x^{2D-4}}\right] \; .
\ee
We can now use identities (\ref{id2}) and (\ref{id3}) to break this into its divergent and finite pieces. We find
\bea
F_{\rm ex,div}(x;x')&=&\f{D(D^2-D-4)}{8(D-1)(D-2)(D-3)}\f{\k^2}{16\pi^D}\G\left(\f{D}{2}\!-\!1\right)\left[\f{1}{a^2}\del^2+2\f{H}{a}\del_0\right]\times \nonumber \\
&&\hspace{2in}\left[\f{i4\pi^{D/2}}{(D-4)}\mu^{D-4}\d^D(x-x')\right] \; , \\
&& \nonumber \\
F_{\rm ex, finite}(x;x')&=&-\f{\k^2}{48\pi^4}\left(\nabla^2+\del_0\del'_0\right)\times\left\{\f{1}{aa'}\del^2\left[\f{\ln(\mu^2\D x^2)}{\D x^2}\right]\right\} \; ,
\eea
where in deriving the divergent part we have used the fact that, in conjunction with the delta function, $\del_0\del'_0\f{1}{aa'}\rightarrow 2\f{H}{a}\del_0-\f{1}{a^2}\del_0^2$. Renormalizing all other terms will follow a very similar procedure. 

Once all of the terms have been renormalized and combined we finally arrive at the full result for the structure functions coming from the 3-point diagram, for convenience we split the results into their finite and divergent pieces
\bea
\lefteqn{F_{3,\rm div}(x;x')=\f{\k^2\G\left(\f{D}{2}\!-\!1\right)\mu^{D-4}}{32\pi^{D/2}(D-4)}\left\{-\f{2(D-1)(D-2)}{(D-3)}H^2\right.} \nonumber \\ 
&&\hspace{1in}\left.+\f{D}{(D-1)}\left[\f{1}{a^2}\del^2+2\f{H}{a}\del_0\right]\right\}i\d^D(x-x') \; , \label{f3div} \\
&& \nonumber \\
\lefteqn{F_{3,\rm finite}(x;x')=-\f{\k^2}{192\pi^4aa'}\del^4\left[\f{\ln(\mu^2\D x^2)}{\D x^2}\right]} \nonumber \\
&&\hspace{1in}+\f{\k^2H^2}{16\pi^4}\Bigg\{\f{3}{4}\del^2\left[\f{\ln(\mu^2\D x^2)}{\D x^2}\right]-\f{1}{2}(\del^2-2\del_0^2)\left[\f{\ln(\f{1}{4}H^2\D x^2)}{\D x^2}\right]+2\del_0^2\f{1}{\D x^2} \nonumber \\
&&\hspace{1in}-9\pi^2i\d^4(x-x')\Bigg\} \label{f3fin} \\
&& \nonumber \\
\lefteqn{G_{3,\rm div}(x;x')=-\f{\k^2H^2\G\left(\f{D}{2}\!-\!1\right)\mu^{D-4}}{32\pi^{D/2}}\left\{\f{D(D^3-11D^2+35D-24)}{(D-1)(D-2)(D-4)}\right\}\times} \nonumber \\
&&\hspace{1in}i\d^D(x-x') \; , \label{g3div} \\
&& \nonumber \\
\lefteqn{G_{3,\rm finite}(x;x')=\f{\k^2H^2}{16\pi^4}\Bigg\{\f{1}{6}\del^2\left[\f{\ln(\mu^2\D x^2)}{\D x^2}\right]-\f{1}{2}\del^2\left[\f{\ln(\f{1}{4}H^2\D x^2)}{\D x^2}\right]} \nonumber \\
&&\hspace{1in}-12\pi^2i\d^4(x-x')\Bigg\} \; . \label{g3fin}
\eea
In the flat space limit $H\rightarrow 0, a=a'=1$ these structure functions recover the old result \cite{LW}. We are now ready to combine the 3-point and 4-point contributions to find the full vacuum polarization.

\subsection{Our Full Result}

We will now find the appropriate counter term coefficients such as to cancel all divergences, and be left with the full renormalized vacuum polarization. Recall that we can actually absorb all of the 4-point contribution, minus the terms proportional to $\ln(a)$, into the counterterms. Thus to find the full counterterm coefficients we simply have to add the divergent coefficients of the 3-point contribution to the the 4-point contribution. From (\ref{f4div}), (\ref{f4fin}), and (\ref{f3div}) we see
\bea
\lefteqn{\overline{C}=\f{\k^2}{16}\left\{\f{(5D^3-25D^2+34D-4)}{8(D-1)(D-3)(D-4)}\times\f{\G\left(\f{D}{2}\!-\!1\right)\mu^{D-4}}{\pi^{D/2}}\right.} \nonumber \\
&&\left.+D(D-5)\times\f{H^{D-4}\G(D-1)\pi}{(4\pi)^{D/2}\G(\f{D}{2})}\cot\left(\f{\pi}{2}D\right)-\f{1}{4\pi^2}\right\} \; ,
\eea
from (\ref{f3div}) we find
\be
C_4=\f{D}{(D-4)(D-1)}\f{\k^2\G\left(\f{D}{2}-1\right)}{128\pi^{D/2}}\mu^{D-4} \; ,
\ee
and from (\ref{g4div}), (\ref{g4fin}), and (\ref{g3div}) we find
\bea
\lefteqn{\D C=\f{\k^2}{16}\left\{\f{D(D^2-6D+3)}{8(D-1)(D-2)}\times\f{\G\left(\f{D}{2}\!-\!1\right)\mu^{D-4}}{\pi^{D/2}}\right.} \nonumber \\
&&\left.-\f{(D^2-4D+1)}{(D-3)}\times\f{4H^{D-4}\G(D-1)\pi}{(4\pi)^{D/2}\G(\f{D}{2})}\cot\left(\f{\pi}{2}D\right)+\f{1}{4\pi^2}\right\} \; .
\eea
Using these coefficients in the counterterms (\ref{DeltaF}) and (\ref{DeltaG}) we can remove all divergences. 
We can also convert each of the three renormalization scales $\mu$ into $\frac12 H$ plus arbitrary finite 
counterterms (denoted by the parameters $\alpha$, $\beta$ and $\gamma$) \cite{GMPW}
\bea
\lefteqn{F(x;x')=\f{\k^2}{8\pi^2}\Bigg\{H^2\left[\ln(a)+\a\right]+\f{1}{a^2}\left[-\f{1}{3}\ln(a)+\b\right]\left(\del^2+2Ha\del_0\right)+\f{H}{3a}\del_0\Bigg\}i\d^4(x-x')} \nonumber \\
&&-\f{\k^2}{1536\pi^4}\f{1}{a}\del^6\Bigg\{\f{1}{a'}\left[\ln^2\left(\f{1}{4}H^2\D x^2\right)-2\ln\left(\f{1}{4}H^2\D x^2\right)\right]\Bigg\}\nonumber \\
&&+\f{\k^2H^2}{128\pi^4}\Bigg\{\left[\f{1}{4}\del^4 + \del^2\del_0^2\right]\ln^2\left(\f{1}{4}H^2\D x^2\right)+\left[-\f{1}{2}\del^4 + 2\del^2\del_0^2\right]\ln\left(\f{1}{4}H^2\D x^2\right)\Bigg\} \; , \label{finalf} \\
&& \nonumber \\
\lefteqn{G(x;x')=\f{\k^2H^2}{6\pi^2}\left[-\ln(a)+\g\right]i\d^4(x-x')} \nonumber \\
&&\hspace{2in}-\f{\k^2H^2}{384\pi^4}\del^4\Bigg\{\ln^2\left(\f{1}{4}H^2 \D x^2\right)-2\ln\left(\f{1}{4}H^2 \D x^2\right)\Bigg\} \; . \label{finalg}
\eea
These structure functions comprise our main result. 

\section{Hartree Approximation}\label{Hartree}

Solving the quantum-corrected Maxwell equation (\ref{QMax}) is an
exercise of comparable difficulty to the one we have just concluded,
so it will appear in a separate publication. In the meantime, we
can gain a qualitative understanding of what the result might
show by applying the Hartree approximation \cite{Howie}. This has
been used previously to study the effect of charged inflationary
scalars on photons \cite{DDPT} and (especially relevant to the
current problem) the effect of inflationary gravitons on massless
fermions \cite{MW1}. In each of the previous cases the Hartree
approximation gave the correct spacetime dependence of the one loop
correction to the mode functions and the correct sign relative to
the tree order result.

The Hartree approximation to (\ref{QMax}) consists of replacing the
Heisenberg operator field equation by its expectation value in free
graviton vacuum and then expanding in terms of coincident graviton
propagators,
\begin{eqnarray}
\lefteqn{0 = \partial_{\mu} \Biggl\{ \sqrt{-g(x)} \, g^{\mu\rho}(x)
g^{\nu \sigma}(x) F_{\rho\sigma}(x) \Biggr\} \; , } \\
& & \hspace{-.5cm} \longrightarrow 0 = \partial_{\mu} \Biggl\{
a^{D-4} \Bigl\langle \Omega_h \Bigl\vert \sqrt{-\widetilde{g}(x)} \,
\widetilde{g}^{\mu\rho}(x) \widetilde{g}^{\nu\sigma}(x) \Bigr\vert
\Omega_h \Bigr\rangle F_{\rho\sigma}(x) \Biggr\} \; , \qquad \\
& & = \partial_{\mu} \Bigl\{ a^{D-4} F^{\mu\nu} \Bigr\} \!+\!
\frac{\kappa^2}{2} \partial_{\mu} \Biggl\{
U^{\mu\nu\rho\sigma\alpha\beta\gamma\delta} a^{D-4}
i\Bigl[\mbox{}_{\alpha\beta} \Delta_{\gamma\delta}\Bigr](x;x)
F_{\rho\sigma}(x) \Biggr\} \!+\! \dots \qquad
\end{eqnarray}
Here $U^{\mu\nu\rho\sigma\alpha\beta\gamma\delta}$ denotes the
tensor factor of the 2-graviton-2-photon vertex, given in expression
(\ref{Uvert}). Most of the coincident graviton propagator is a
divergent constant; the secular effects for which we are searching
derive from only the logarithm part of the $A$-type propagator
(\ref{DeltaA}). At this point we can also take $D=4$,\footnote{One
might worry about factors of $\ln(a)$ which could arise when a
divergent constant multiplies $a^{D-4}$. However, we can see from
expressions (\ref{DeltaF}) and (\ref{DeltaG}) that the very same factor
of $a^{D-4}$ multiplies the counterterms which absorb divergences from
the 4-point diagram. So there can be no secular contributions from this
source and we may as well drop the divergent constants and take $D=4$.}
so our Hartree approximation to the effective field equation
(\ref{QMax}) is,
\begin{equation}
0 = \partial_{\mu} F^{\mu\nu} + \frac{\kappa^2 H^2}{8 \pi^2} \,
\partial_{\mu} \Biggl\{ U^{\mu\nu\rho\sigma\alpha\beta\gamma\delta}
\Bigl[\mbox{}_{\alpha\beta} T^A_{\gamma\delta}\Bigr] \times
\ln(a) F_{\rho\sigma} \Biggr\} + O(\kappa^4) \; . \label{ourHart}
\end{equation}
Recall that the $A$-type tensor factor was defined in expression
(\ref{TA}).

Substituting (\ref{Uvert}) and (\ref{TA}) in (\ref{ourHart}) gives a
simple result,
\begin{equation}
0 = \partial_{\mu} F^{\mu\nu} + \frac{\kappa^2 H^2}{4 \pi^2} \,
\partial_{\mu} \Biggl\{ \ln(a) \Bigl[ -3 F^{\mu\nu} \!+\!
4 F^{\overline{\mu} \nu} \!+\! 4 F^{\mu \overline{\nu}} \!-\!
3F^{\overline{\mu} \overline{\nu}} \Bigr] \Biggr\} + O(\kappa^4) \; .
\label{Hart2}
\end{equation}
Here a barred index on any tensor indicates that its 0-component has
been suppressed, for example, $V^{\overline{\mu}} \equiv
\overline{\eta}^{\mu\nu} V_{\nu} = V^{\mu} - \delta^{\mu}_0 V^0$.
We can distinguish in expression (\ref{Hart2}) between the cases
of $\nu = 0$ and $\nu = i$. The constraint equation is effectively
multiplied by a secular factor,
\begin{eqnarray}
0 & = & \partial_j F^{j0} + \frac{\kappa^2 H^2}{4 \pi^2} \, \partial_j
\Biggl\{ \ln(a) F^{j 0} \Biggr\} + O(\kappa^4) \; , \qquad \\
& = & \Biggl\{ 1 + \frac{\kappa^2 H^2}{4 \pi^2} \, \ln(a) \Biggr\}
\partial_j F^{j0} + O(\kappa^4) \; . \label{Hart3a} \qquad
\end{eqnarray}
This has no effect on dynamical photons although it would lead to a
secular screening of a point charge of the sort reported by Kitamoto
and Kitazawa \cite{HKYK} provided there is no compensating secular
factor on the charge density. The equation of relevance for
dynamical photons is $\nu = i$,
\begin{equation}
\partial_{\mu} F^{\mu i} + \frac{\kappa^2 H^2}{4 \pi^2} \Biggl\{
\partial_0 \Bigl[ \ln(a) F^{0i} \Bigr] \!+\! \partial_j \Bigl[2 \ln(a)
F^{ji} \Bigr] \Biggr\} + O(\kappa^4) \; . \label{Hart3b}
\end{equation}

The one loop correction to the effective field equation of course
fixes only the one loop corrections to the field strength. We
therefore expand in powers of $\kappa^2$,
\begin{equation}
F^{\mu\nu} = F^{\mu\nu}_{(0)} + \kappa^2 F^{\mu\nu}_{(1)} + \kappa^4
F^{\mu\nu}_{(2)} + \dots
\end{equation}
Equations (\ref{Hart3a}) and (\ref{Hart3b}) imply the following relations
for the one loop field strengths,
\begin{eqnarray}
\partial_j \kappa^2 F^{j0}_{(1)} & = & 0 \; , \label{Hart4a} \\
\partial_{\mu} \kappa^2 F^{\mu i}_{(1)} & = & -\frac{\kappa^2 H^2}{4 \pi^2}
\Biggl\{ \partial_0 \Bigl[ \ln(a) F^{0i}_{(0)} \Bigr] \!+\! \partial_j
\Bigl[2 \ln(a) F^{ji}_{(0)} \Bigr] \Biggr\} \; . \qquad \label{Hart4b}
\end{eqnarray}
With the $U(1)$ Bianchi identity, the leading secular behavior is,
\begin{eqnarray}
\kappa^2 F^{0i}_{(1)} & \longrightarrow & -\frac{\kappa^2 H^2}{4 \pi^2}
\, \ln(a) \times F^{0i}_{(0)} \; , \label{Hart5a} \\
\kappa^2 F^{ij}_{(1)} & \longrightarrow & -\frac{\kappa^2 H^2}{4 \pi^2}
\, \frac{\ln(a)}{H a} \times \Bigl[ \partial^i F^{0j}_{(0)} \!-\!
\partial^j F^{0i}_{(0)} \Bigr] \; . \qquad \label{Hart5b}
\end{eqnarray}
We see that the one loop correction to the electric field strength
of a photon tends to cancel its classical value whereas the one loop
correction to the magnetic field strength dies off.

%--------------------------------------------------------------------------------------------------------------------------------------------------------------%

\section{Discussion}\label{Discuss}

We have used dimensional regularization to compute the one loop quantum 
gravitational contribution to the vacuum polarization on de Sitter 
background. We first calculated the 4-point contribution in section 
\ref{4point}, and then derived the much more cumbersome 3-point 
contribution in section \ref{3point}. Each result was expressed 
in the form (\ref{F+G}) as the sum of two transverse projection 
operators acting on structure functions. In sub-section~\ref{counter}
the relevant BPHZ counterterms (\ref{LBPHZ}) were also reduced to
this form, resulting in expressions (\ref{DeltaF}-\ref{DeltaG}).
Renormalization was implemented in section~\ref{Renorm} to give our
final results (\ref{finalf}) and (\ref{finalg}) for the structure 
functions $F(x;x')$ and $G(x;x')$.

Our ultimate goal is to study how inflationary gravitons affect
electrodynamics using the quantum-corrected Maxwell equation
(\ref{QMax}). Specializing to de Sitter in conformal coordinates, 
substituting our form (\ref{F+G}) for representing the vacuum 
polarization, and partially integrating the primed derivatives, 
allows us to express the quantum-corrected Maxwell equations as,
\begin{equation}
\partial_{\nu} F^{\nu\mu}(x) + \partial_{\nu} \! \int \!\! d^4x' 
\Biggl\{ i F(x;x') F^{\nu\mu}(x') \!+\! i G(x;x') F^{\bar{\nu}
\bar{\mu}}(x') \Biggr\} = J^{\mu} \; . \label{OurMax}
\end{equation}
(Recall that a barred index is purely spatial.) Equation (\ref{OurMax})
can be employed the same way one uses the classical Maxwell equation
to study dynamical photons ($J^{\mu} = 0$ solutions) and the 
electric and magnetic fields induced by standard sources. We have 
already done this for the one loop vacuum polarization from gravitons on
flat space background \cite{LW}. Closely related studies have also
been made of the effects that inflationary scalars have on dynamical
photons \cite{SQED} and on electrodynamic forces \cite{DW}.

The actual implementation of this program requires solving 
integro-differential equations in the context of the Schwinger-Keldysh
formalism \cite{SK}. That is an project comparable to the one we 
have just completed, so it will be deferred to a separate work. 
However, a simple estimate of what it might give was derived in 
section \ref{Hartree} by making the Hartree approximation \cite{Howie}
to localize the effective field equation. We find that the one loop
electric field strength (\ref{Hart5a}) of dynamical photons experiences 
a secular growth which tends to cancel its free field value, whereas 
the one loop correction to the magnetic field strength (\ref{Hart5b}) 
dies away compared to its free field value. It is interesting to note 
that the magnetic response to inflationary scalars also seems to be 
subdominant to the electric repsonse \cite{DW}.

Working out what the full equation (\ref{OurMax}) gives for dynamical
photons is important to check the observation in \cite{MW2} that the
spin-spin interaction between gravitons and fermions seems to explain
why inflationary gravitons cause the fermion mode function to grow
\cite{MW1} whereas they have no secular effect on the mode function of
a massless, minimally coupled scalar \cite{KW}. Another important
exercise is to work out the effect of inflationary gravitons on the 
electric field of a point charge. This is the natural way to check the 
surprising claim of Kitamoto and Kitazawa that infrared gravitons screen 
sub-horizon interactions during inflation \cite{HKYK}.

Before closing, we should comment on the gauge issue. The vacuum 
polarization requires fixing both the $U(1)$ and diffeomorphism 
symmetries, and the manner in which this is accomplished can affect 
the result. Our previous study of gravitons on flat background 
revealed no dependence upon the choice of electromagnetic gauge, 
but a huge variation with the gravitational gauge \cite{LW}. We 
believe there is not likely to be any gauge dependence in the
leading secular infrared effects one finds from de Sitter gravitons 
because the spin two part of the graviton propagator has the same 
infrared logarithm term in any gauge \cite{KMW,MTW5}. Note that it 
is perfectly possible for a 1PI function such as the vacuum polarization 
to change with the gauge, while a particular feature of its dependence on 
space and time is the same in all gauges \cite{MW3}. That is precisely 
what happens with the pole terms of 1PI functions in flat space quantum 
field theory, and we suspect that the same applies for the leading 
secular dependence on de Sitter.

\vskip .5cm

\centerline{\bf Acknowledgements}

We are grateful to S. Deser for having suggested that we make
this computation. This work was partially supported by NSF grant
PHY-1205591 and by the Institute for Fundamental Theory at the
University of Florida.

\end{document}